\documentclass[twocolumn,pra,aps,superscriptaddress,showpacs]{revtex4-1}

\usepackage{mathptmx}
\usepackage{dcolumn}
\usepackage{amsmath,amssymb,amsfonts}
\usepackage{dsfont}
\usepackage{bm}
\usepackage{color}
\usepackage{latexsym}
\usepackage{epstopdf}
\usepackage[english]{babel}
\usepackage{enumerate}

\usepackage[pdftex]{graphicx}

\usepackage{natbib}
\usepackage{epstopdf}
\usepackage{appendix}

\definecolor{mygrey}{gray}{0.35}
\definecolor{myblue}{rgb}{0.2,0.2,0.8}
\definecolor{myzard}{cmyk}{0,0,0.05,0}
\definecolor{mywhite}{rgb}{1,1,1}
\definecolor{mywhite}{rgb}{1,1,1}
\definecolor{myred}{rgb}{1,0.,0.3}

\usepackage[colorlinks=true,citecolor=myblue,linkcolor=myred]{hyperref}

\def\be{\begin{equation}}
\def\ee{\end{equation}}
\def\ba{\begin{align}}
\def\enda{\end{align}}
\def\bi{\begin{itemize}}
\def\ei{\end{itemize}}

 \def\ee{\mathord{\rm e}}

\renewcommand{\ee}{{\rm e}}
\renewcommand{\aa}{{\rm a}}

\def\beq{\begin{equation}}
\def\eeq{\end{equation}}

 \newcommand{\ket}[1]{|#1\rangle}
 \newcommand{\bra}[1]{\langle #1|}

 \newcommand{\ketbradif}[2]{\ket{#1}\bra{#2}}
 \newcommand{\ketbra}[1]{\ketbradif {#1}{#1}}

\def\oned{\mathrm{1d}}

\newcommand{\braket}[2]{\langle #1|#2\rangle}

\newcommand{\bb}{\mathrm{b}}
\newcommand{\rt}{\mathrm{t}}
\newcommand{\rs}{\mathrm{s}}
\newcommand{\rd}{\mathrm{d}}
\newcommand{\rg}{\mathrm{g}}

\begin{document}
	
\title{Reliable multiphoton generation in waveguide QED}

 \author{A. Gonz\'{a}lez-Tudela}
 \affiliation{Max-Planck-Institut f\"{u}r Quantenoptik Hans-Kopfermann-Str. 1.
85748 Garching, Germany }

 \author{V. Paulisch}
 \affiliation{Max-Planck-Institut f\"{u}r Quantenoptik Hans-Kopfermann-Str. 1.
85748 Garching, Germany }

 \author{H. J. Kimble}
  \affiliation{Max-Planck-Institut f\"{u}r Quantenoptik Hans-Kopfermann-Str. 1.
85748 Garching, Germany }
 \affiliation{Norman Bridge Laboratory of Physics 12-33}
  \affiliation{Institute for Quantum Information and Matter, California Institute of Technology, Pasadena, CA 91125, USA}

\author{J. I. Cirac}
 \affiliation{Max-Planck-Institut f\"{u}r Quantenoptik Hans-Kopfermann-Str. 1.
85748 Garching, Germany }
\date{\today}

\begin{abstract}
In spite of decades of effort, it has not yet been possible to create single-mode multiphoton states of light with high success probability and near unity fidelity. Complex quantum states of propagating optical photons would be an enabling resource for diverse protocols in quantum information science, including for interconnecting quantum nodes in quantum networks. Here, we propose several methods to generate heralded mutipartite entangled atomic and photonic states by using the strong and long-range dissipative couplings between atoms emerging in waveguide QED setups. Our theoretical analysis demonstrates high success probabilities and fidelities are possible exploiting waveguide QED properties.
\end{abstract}

\maketitle

On-demand generation of optical propagating photons is
at the basis of many applications in quantum information science, including multipartite teleportation \cite{murao99a}, quantum repeaters \cite{briegel98a}, quantum cryptography \cite{gisin02a,durkin02a}, and quantum metrology \cite{giovannetti04a}. While single photons are routinely produced in different experimental setups \cite{lounis05a}, {\em single--mode multiphoton states} are much harder to generate \cite{dellanno06a}. Current methods are
limited by either an exponentially small success probabilities or by low fidelities. Here we show that reliable single-mode multiphoton sources can be constructed with emitters close to nano-photonic waveguides by  combining both the strong coupling present in those systems and collective phenomena. In particular, we show how collective excitations can be efficiently loaded in a collection of atoms, which can then be released with a fast laser pulse to produce the desired multi-photon state \cite{porras08a,gonzaleztudela15a}. In order to boost the success probability and fidelity of each excitation process, our method utilizes atoms to both generate the excitations in the rest, as well as to herald the successful generation. Furthermore, to overcome the exponential scaling of the probability of success with the number of excitations, we design a protocol to merge excitations that are present in different internal atomic levels with a polynomial scaling.

The enhancement of light-matter interactions provided by quantum nanophotonics opens up new avenues to create high-fidelity multiphoton states. For example,  $m$ quantum emitters can be strongly coupled to structured waveguides, which show large Purcell factors, $P_{\oned}$, so that $m$ atomic excitations can be mapped to a waveguide mode with an error (or infidelity, $I_m$) scaling as $m/P_\oned$. However, the resulting state is not a single mode, but a complex entangled state of several modes \cite{gonzaleztudela15a}, so that it cannot be directly used for quantum information purposes. Single-mode multiphoton states can be created by storing $m$ collective excitations in $N\gg m$ atoms, which are then mapped to a photonic state of the waveguide. While the latter process can be achieved with very low infidelity, scaling as $m^2/( N P_\oned)$ \cite{porras08a,gonzaleztudela15a}, present schemes for the first part scale like $I_m\propto m/\sqrt{P_{\oned}}$ \cite{gonzaleztudela15a}, as they still do not 
fully exploit the strong 
coupling to the waveguide nor collective effects. This ultimately limits the fidelity of the whole procedure.

In this work we show how this limitation can be overcome with new schemes for the heralded generation of $m$ collective excitations in $N\gg m$ atoms coupled to a waveguide. The basic idea is to use the atoms themselves to both create the excitations one-by-one, and to herald the success of the process. In this way, arbitrarily small infidelities, $I_m$, can be obtained at the expense of making the process non-deterministic. Depending on the scheme, we find that the global probability of success (or, inversely, the average number of operations, $R_m$) decreases (increases) exponentially with $m$, and thus it cannot be scaled to arbitrarily large photon production. Finally, we also show how to overcome this exponential law by using additional atomic states, atom number resolved detection, and a specific protocol to merge excitations, while keeping a very low global infidelity, $I_m \sim {\rm poly}(m)/(NP_\oned)$.

\begin{figure}[tb]
	\centering
	\includegraphics[width=0.44\textwidth]{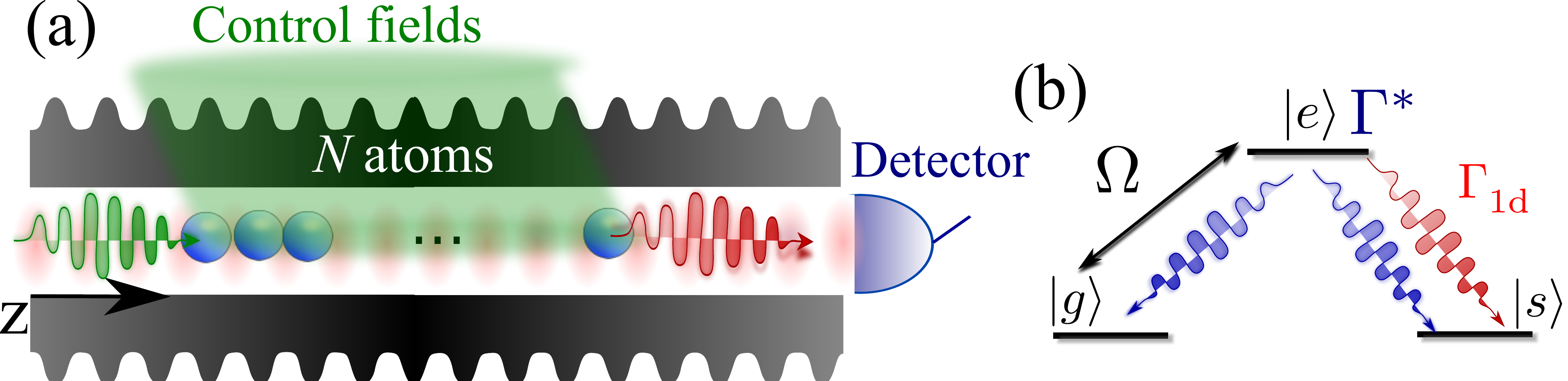}
	\caption{(a) Setup for the first protocol: $N$ \emph{target} atoms are collectively coupled to the waveguide. A photon detection heralds the addition of a collective excitation. (b) Atomic level structure: waveguide modes are coupled to the $e\leftrightarrow s$ transition with a spontaneous emission rate $\Gamma_\oned$. The transition $g\leftrightarrow e$ is driven by a Raman laser $\Omega$ and the spontaneous emission rate to other modes is denoted by $\Gamma^*$.}
	\label{fig1}
\end{figure}

\begin{figure*}[tb]
	\centering
	\includegraphics[width=0.99\textwidth]{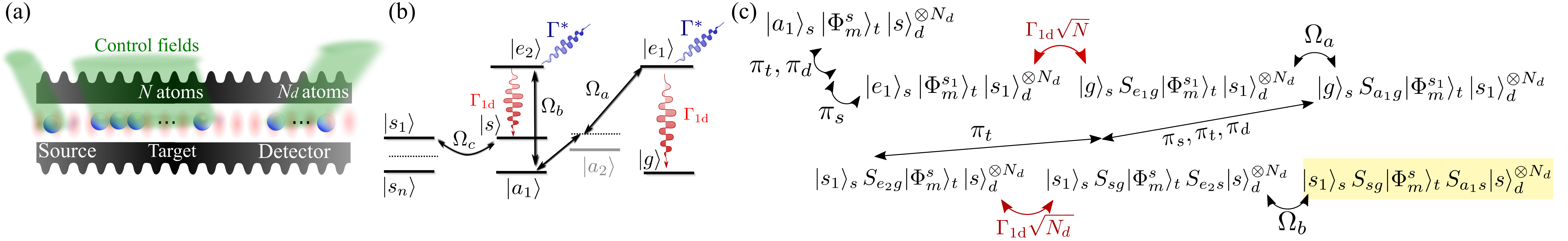}
	\caption{(a) Setup for the second and third protocol: $N$ {\em target}, one {\em source} and $N_d$ \emph{detector} atoms are collectively coupled to the waveguide. The source atom creates a single excitation, and the detector atoms herald the success of the process. (b) Atomic level structure: waveguide modes are coupled to the $e_1\leftrightarrow g$ and $e_2\leftrightarrow s$ transition with a spontaneous emission rate $\Gamma_\oned$. A two-photon transition $a_1\leftrightarrow e_1$ is driven by laser light via level $a_2$ with effective Rabi frequency $\Omega_a$. Transition $a_1\leftrightarrow e_2$ is driven by a different laser with Rabi frequency $\Omega_b$. Coupling between levels $s,s_1$ is characterized by a Rabi frequency $\Omega_c$. Other levels $\{s_n\}$ can be manipulated using lasers/microwave fields. (c) Scheme for the second protocol, with $\pi_{c,t,d}$ denoting the $\pi$-pulses for the population transfers within the source/target/detector atoms. The driven transitions to the excited states 
are indicated by the Rabi frequencies $\Omega_{a,b}$. In shaded yellow the final state where we want to arrive is indicated.}\label{fig2}
\end{figure*}

Structured waveguide setups with trapped atoms offer several interesting characteristics that we exploit to design our protocols, namely, i) regions of large Purcell Factor $P_\oned \gg 1$, e.g., due to slow light in engineered dielectrics \cite{laucht12a,lodahl15a,yu14a,goban15a}, while keeping at the same time ii) long-propagation lengths of the guided modes compared to the characteristic wavelength ($\lambda_\aa$) that give rise to long-range dissipative couplings \cite{chang12a}.  Moreover, as shown in, e.g. Ref.~\cite{chang12a}, in order to avoid dipole-dipole interactions and fully exploit superradiance effects we assume iii) the atoms to be placed at distances $z_n=n \lambda_\aa$, with $n\in \mathbb{N}$. \footnote{In fact all results can be extended to separations of $\lambda_\aa/2$}. Finally, iv) it is possible to read the collective atomic state very efficiently through the waveguide due to the naturally short timescales and large collection efficiencies. We use atomic detection for heralding, which 
has already reached accuracies of $10^{-4}$ \cite{myerson08a} in 
trapped ion setups, where the collection efficiency is not enhanced by the presence of a waveguide, as in our scheme. Thus, here we assume the atomic detection to be perfect, and consider $P_\oned$ and $N$ as the main resources to analyze the figures of merit of the protocols. For the analysis of three different schemes, we adopt the following strategy for each one: we analyze the process of generating a single collective excitation, assuming the atoms already store $m$ excitations. We denote by $p$ the probability of success, and $I_{m\rightarrow m+1}$ the corresponding infidelity. We finally analyze the average number of operations $R_m$ and final infidelities $I_m$ for accumulating all $m$ excitations. To simplify the expressions along the main text, we assume to work in a regime with $N\gg m$ and $P_\oned\gg 1$, though complete expressions can be found in Supplementary Material~\cite{supmat}.

Let us first analyze a protocol inspired in a method originally devised to create long-distant entangled states in atomic ensembles \cite{duan01a}, that requires $N$ atoms placed close to a 1d waveguide as depicted in Fig.~\ref{fig1}(a). The atoms must have a level structure as depicted in Fig.~\ref{fig1}(b), where the atomic transition $e\leftrightarrow s$ is coupled to the guided modes at a rate $\Gamma_\oned$ and the transition $g\leftrightarrow e$ is driven equally by a laser with Rabi frequency $\Omega$. The excited states $e$ also radiate into leaky modes (of the waveguide or outside) other than the relevant one, that give rise to a decay rate $\Gamma^*$, leading to the Purcell factor $P_\oned=\Gamma_\oned/\Gamma^*$. The excitations are stored in the states $g$ and $s$, which are assumed to be decoherence-free like any other hyperfine ground state. The procedure assumes that we start with $m$ collective excitations in level $s$, and the goal is to add another one. Thus, the 
initial state is $\ket{\Phi_m^s}\propto\mathrm{sym}\{\ket{s}^{\otimes m}\otimes \ket{g}^{N-m}\}$, where `sym' denotes the symmetrizing operator. The idea is to weakly excite the atoms in $g$ to level $e$ with a short laser pulse of duration $T\ll 1/(N\Gamma_\oned)$ and if a photon in the waveguide is detected, it heralds the addition of an excitation in state $s$. As all the atoms are equally coupled to the waveguide, the excitation will be collective. 

Let us denote by $x=\Omega \sqrt{N}T/2\ll 1$, so that right after the pulse we have the state (up to a normalization constant), $\big[\mathbf{1}+x\,S_{eg}+x^2\, S_{eg}^2+O(x^3)\big]\ket{\Phi_m^s}$, where we used the notation $S_{\alpha\beta} = (N)^{-1/2}\sum_{n=1}^N \sigma_{\alpha\beta}^{n}$ for the collective spin operators and $\sigma^{n}_{\alpha\beta}=\ket{\alpha}_n\bra{\beta}$ with atom number $n=1,\dots,N$. After the pulse, we leave the system free to evolve under the interaction with the waveguide for a time $t \gg 1/\Gamma^*)$, in which case the wavefunction terms with excitations in $e$ decay either to waveguide/leaky photons. If a waveguide photon is detected, either it comes from the lowest order term, $O(x)$, in which case the atomic state will be the desired one, i.e., $S_{sg}\ket{\Phi_m^s}$, or from the double excited state, in which case we will prepare the wrong state introducing an error. Emission of leaky photons also produces errors, but of smaller order \cite{supmat}. Denoting by $\eta$ 
the detection efficiency, the success probability in generating the desired state is $p\approx \eta x^2$ and the infidelity is $I_{m\rightarrow m+1} \propto (1-\eta)x^2$ [to lowest order in $x$]. To create $m$ excitations, one has to detect $m$ consecutive photons, leading to $R_m=p^{-m}$ and $I_m\propto  m(1-\eta)x^2$. As in Ref.~\cite{duan01a}, the error can be made arbitrarily small at the expense of decreasing the success probability. If a high fidelity is required (e.g., $0.999$) the method is practicable only for few excitations. One way of reducing $R_m$ is by using an additional metastable state $s_1$ in which the heralded excitation is stored after each successful addition. This can be done by combining, e.g., a two-photon Raman transition, to make a beam splitter transformation between levels $s$ and $s_1$, and post-selection conditioned on no atomic detection in $s$. Then, assuming that we have $m$ excitations already stored in $s_1$, this procedure generates $m+1$ within the same level. It can be 
shown \cite{supmat,dakna99a} that this strategy leads to an average number of operations $R_m\propto e^m/p$.

To overcome the trade-off between probabilities and fidelities coming from zero and double excitations, we propose a protocol relying on a configuration as depicted in Fig.~\ref{fig2}(a): we replace the write field of the previous scheme by a single \emph{source} atom that guarantees the transfer of at most a single excitation to the \emph{target} ensemble. Furthermore, in a second step, $N_d$ \emph{detector} atoms act as a detector to herald the transfer of excitations, replacing the photon detector. Both the source and detector atoms should be separated from the target ensemble by a multiple of $\lambda_\aa/2$ and sufficiently separated for independent addressing with external control fields.
The protocol requires a level structure as shown in Fig.~\ref{fig2}(b) where two dipolar transitions $e_1\leftrightarrow g$ and $e_2\leftrightarrow s$ are coupled to the same waveguide mode with rates that we set to be equal for simplicity and with a corresponding Purcell Factor $P_\oned$. We require the use of other hyperfine, auxiliary levels, $a_1,a_2,s_1$. The transition $a_1\leftrightarrow e_2$ is connected by a laser, whereas the $a_1\leftrightarrow e_1$ is a two-photon transition mediated by $a_2$, with effective Rabi frequency $\Omega_a$, so that direct spontaneous emission from $e_1\to a_1$ is forbidden \cite{vanenk97a,porras08a,borregaard15a,supmat}. The level $s_1$ is used to store excitations and decouple them from the dynamics induced by the waveguide. We require that $s_1$ is not connected to neither $e_1$ nor $e_2$ by a dipole transition, so that it will only be connected to $s$ coherently through microwave fields.

As before, this protocol starts with a superposition in the target ensemble $\ket{\Phi_m^{s}}_t$, and with the source/detector atoms in $a_1/s$ respectively. The heralded transfer of a single collective excitation consists of several steps [Fig.~\ref{fig2}(c)]. The first one, coherently transfers the excitations from $s\to s_1$ in the target and detector ensemble to protect them from the waveguide dynamics. The second step uses a short laser pulse in the source atom to excite it to state $e_1$, and then switches on the lasers driving $e_1\leftrightarrow a_1$  via a two-photon transition in the target ensemble with (effective) Rabi frequency $\Omega_a$ for a time $T_a$. Ideally, the source atom exchanges the excitation with the target ensemble by a virtual photon in the waveguide, thus generating a collective excitation in $a_1$-state of the target ensemble. After that, the laser $\Omega_a$ is turned off and one waits for a time $t\gg \left(\Gamma^*\right)^{-1}$ such that any (non-ideal) remaining population 
in the 
excited state decays. Thirdly, we apply $\pi$-pulses to decouple the source atom and couple the target and detector ensemble. Another short pulse is applied to move the collective excitation in the target ensemble from $a_1$ to $e_2$ and we switch on the laser $\Omega_b$ driving the $e_2\leftrightarrow a_1$ transition in the detector ensemble for a time $T_b$ with Rabi frequency $\Omega_b$. This transfers the collective excitation in the target ensemble to $s$ and creates a collective excitation in $a_1$ in the detector ensemble. At the end, a measurement of that internal state, $a_1$, of the detector atoms (through fluorescence in the waveguide) heralds the successful preparation of a collective excitation in the target ensemble, i.e., $\ket{\Phi_{m+1}^{s}}_t$.

\begin{figure}[tb]
	\centering
	\includegraphics[width=0.45\textwidth]{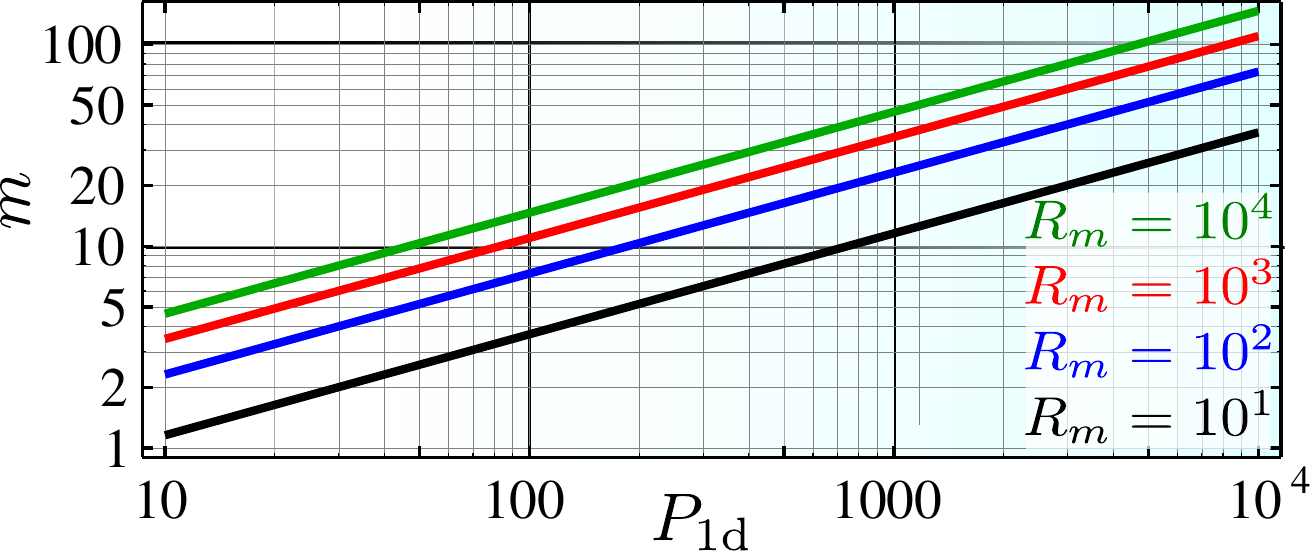}
	\caption{Approximate number $m$ of collective excitations with $I_m=0$ for a fixed number of repetitions $R_m$ (see legend) for the method where we directly accumulate excitations such that $R_m=p^{-m}$ and $p\approx p_a^2$, with $p_a$ as defined in the text and assuming $N,N_d\gg m$.}\label{fig3}
\end{figure}

Let us now analyze the protocol in detail. In the second step, the evolution of the source/target atoms is described by a master equation \cite{supmat}, which can be analytically solved in the limit $N P_\oned \gg 1$. By choosing the Rabi frequency $\Omega_a\approx \sqrt{N \Gamma_{\oned }\Gamma^*}$, and $T_a=\pi/\sqrt{\Gamma_\oned \Gamma^*}$, the probability for the ensemble to end up in the desired state after the second step, $\ket{g}_s\otimes S_{a_1g}\ket{\Phi_{m}^{s_1}}_t$, is maximized $p_\aa\approx e^{-\pi/\sqrt{P_{\oned}}}$, which can be improved by appropriate pulse shaping \cite{supmat}. Similarly, in the third step, the evolution of the source/detector atoms can be analytically solved in the limit $N_d P_\oned\gg 1$, obtaining a probability to end up in the desired state $S_{s g}\ket{\Phi_{m}^{s}}_t\otimes S_{a_1 s}\ket{s}_d^{\otimes N_d}$ given also by $p_\bb \geq p_\aa$. Thus, the total probability of success of the protocol is lower bounded by $p\gtrsim p_\aa^2$, with a heralding infidelity $I_{
m\to m+1}=0$, as we rule out all the possible errors through post-selection. Indeed, the only way to have the detector atom in $a_1$ is that the steps two and three have occurred as desired. Any spontaneous emission in any of the atoms or photon absorption is incompatible with that event.
Thus, to accumulate $m$ excitations we need to repeat an average number of times $R_m\propto p^{-m}$, but with $I_{m}=0$. As $p$ can be very close to $1$ for systems with $N,N_d,P_\oned\gg 1$, we can still expect to achieve big number of excitations $m\gg1$ in spite of the exponential scaling of $R_m$ as shown in Fig.~\ref{fig3} for systems with $N,N_d\gg m$. In case of unsuccessful detection, then, we pump all the target atoms back to $\ket{g}$ and re-initialize the process.

\begin{figure}[tb]
	\centering
	\includegraphics[width=0.5\textwidth]{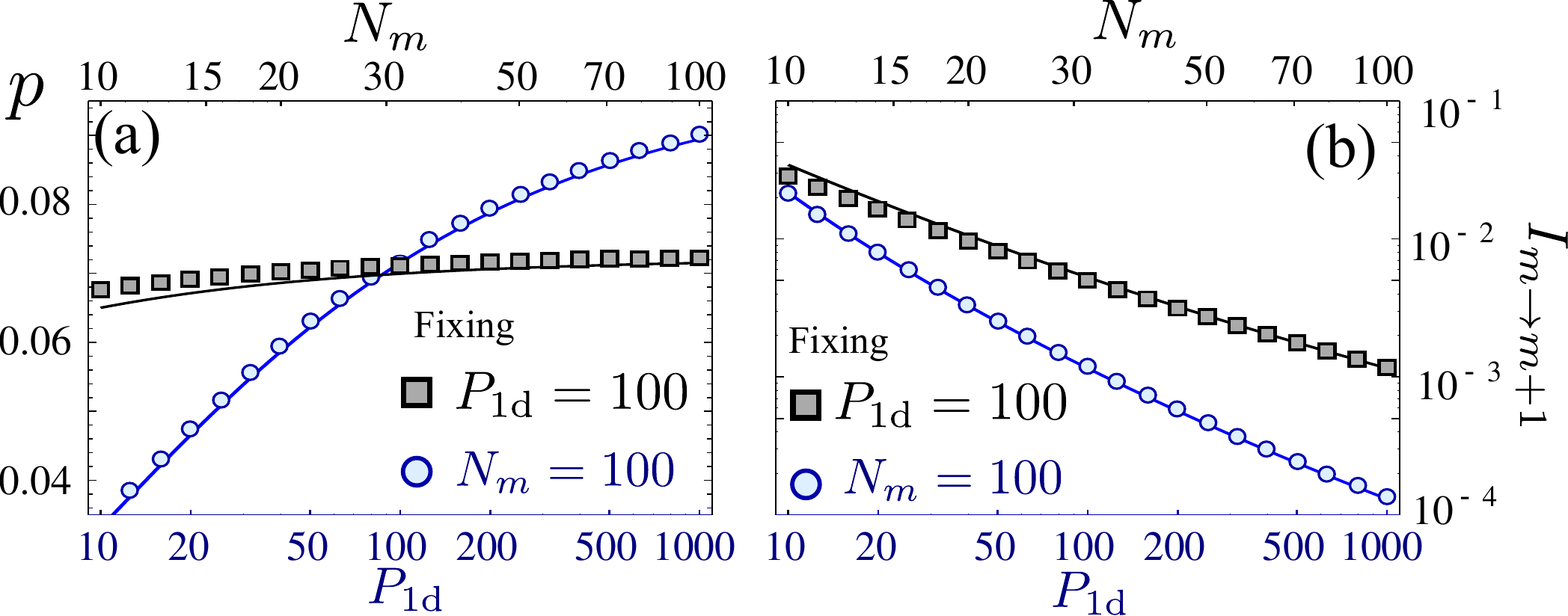}
	\caption{Numerical calculation (markers) and analytical approximation (solid lines) of total probability $p$ and $I_{m\to m+1}$ for the third protocol discussed in the text in panel (a) and (b) respectively. We plot the variation with $P_\oned$ for fixed $N_m=100$ (in blue) and with $N_m$ for fixed $P_\oned=100$. The analytical approximations for $p_a, p_b$ are given in the text and the analysis is made in Ref.~\cite{supmat}. The curves are obtained from full numerical integration of the master equation.}\label{fig4}
\end{figure}

After this analysis one natural question arises, that is, whether the exponential number of operations is a fundamental problem of probabilistic protocols \cite{dakna99a}. This leads to our third and final scheme in which we design a specific protocol which circumvents this exponential scaling with a judicious modification of the previous protocol using several additional atomic states $s_{n}$ [see Fig.~\ref{fig2}(b)] and atomic number resolved detection. The idea is that with each successful detection, we transfer the heralded collective excitation in $s$ not to the same level $s_1$, but to one of several states $\{s_n\}$ to then \emph{merge} them a posteriori with an adequate protocol that we explain below. 
For that note that, in contrast to our previous schemes, in case of unsuccessful detection, i.e., the detector atoms were not found in $a_1$, the $m$ excitations stored in $\{s_n\}_n$ are not destroyed, but only the one we wanted to add. We can pump back the target atoms in $s,a_1$ to $g$ and try again. The price one has to pay is that errors appear since one may not recover the original collective state in the ensemble, i.e., because a spontaneous emission event occurred within the target ensemble \cite{supmat}. One can show that the main source of errors occurs 
when a collective excitation was indeed produced in the ensemble, but it was not detected (because, e.g., of spontaneous emission in the detector ensemble). In order to reduce this error, we have to ensure that undetected collective states return back to $g$ coherently. For that, we apply the following repumping procedure:  first, we pump all levels but $s,\{s_n\}_n$ back to $g$. Then we coherently transfer the potential excitations from $s\to a_1$, and then apply a pumping through the waveguide, $a_1\to e_1\to g$. 
In order to further minimize the errors introduced by repumping and obtain the desired scaling of infidelity ($1/(N P_\oned)$), we also need to modify the third step of the protocol for which then, interestingly, only a single detector atom is needed, $N_d=1$. It can be shown that once we are in $S_{a_1 g}\ket{\Phi_m^{\{s_n\}_n}}\ket{s}_d$, the optimal strategy is to apply a fast $\pi$-pulse in the target atoms such $S_{e_2 g}\ket{\Phi_m^{\{s_n\}_n}}_t$ is prepared. Then, the system is left free to interact for a time $T_\bb=1/\Gamma_\oned$ without any external fields, and the dynamics are terminated by applying another $\pi$-pulse on the target and detector atoms. We find that the optimal probability for the system to be in $S_{s g}\ket{\Phi_{m}^{\{s_n\}_n}}_t \otimes \ket{a_1}_d$ is $p_\bb \approx 0.1 \left(1-1/P_\oned\right)$, which can be improved up to $p_\bb\sim 0.33$ by repeating the fast $\pi$-pulses several times \cite{supmat}. Finally, if we fail, we apply the repumping and repeat the process until 
success, of the order of $1/p$ times with $p=p_\aa p_\bb$. In Ref.~\cite{supmat}, we analyze the process in detail, identifying all errors and how they accumulate in the repetitions, leading to an (averaged) infidelity $I_{m\to m+1}\lesssim m/(p N P_\oned)$ \cite{supmat}. We confirm numerically (Fig.~\ref{fig4}) the scaling of $p$ and $I_{m\to m+1}$ integrating the corresponding master equation, showing that our analytical estimations capture the right scalings.

Thus, the problem is reduced to \emph{merging} the atomic excitations distributed in the levels $\{s_n\}_n$. For example, if instead of adding excitations one by one, as we did for the first scheme, we use $\log_2 m$ metastable states $s_n$, and adopt a tree-like structure, we obtain a superpolynomial scaling \cite{fiurasek05a,supmat} in $R_m\propto m^{\log_2 m}/p$ to merge $m$ collective excitations in a single atomic level. The key step is to double the number of excitations in each step using beam splitter transformations and post-selection conditioned no atomic detection in one of the levels. Moreover, combining the beam splitters with displacement transformations, arbitrary superpositions of collective atomic excitations can be obtained~\cite{dakna99a,fiurasek05a,supmat}.

Finally, we go one step further by using number-resolved atomic detection to obtain a polynomial scaling. The key point is to realize that if we have $n$ excitations stored in two atomic levels, after applying a beam splitter transformation, the probability to obtain exactly $2n$ excitations in one of them (by detecting no excitation in the other) decays with $n$; however, the probability of obtaining more than $3n/2$ in one level (by detecting $p<n/2$ in the other) is actually bounded below by 1/3, independent of $n$ \cite{supmat}. Assuming the worst case scenario in which after detecting $q<m/2$ excitations in one of the two levels, we assume that the other state only goes up to $3n/2$ excitations, that gives us an upper bound for the number of operations \cite{supmat} $R_{m}\lesssim m^{4.41}\log_{3/2} (m)/p$, that is already polynomial, and which in practical situations is better as the number of excitations in the other mode is always larger than $3n/2$. The number of atomic levels, $\ket{s_i}$, required 
to reach $m$ excitations scales logarithmically $\log_{3/2} m$, and the final infidelity scales as $I_m\propto \mathrm{poly}(m)/(N P_\oned)$. We note that atomic detection may itself introduce errors, however, the aim of this scheme is just to show that polynomial scaling can be achieved, despite not being currently the most efficient method. However, we expect that by suitable modification and adaptation to specific setups, polynomial and efficient scaling can be constructed.

State-of-the art technologies already provide systems with $N\sim 1-2$ emitters coupled to engineered dielectrics \cite{thompson13a,tiecke14a,yu14a,goban15a,laucht12a,lodahl15a} with $P_\oned\sim 1-100$. Advances in both fabrication and trapping techniques foresee implementations with $N\sim 100$ atoms and $P_\oned\gtrsim 100$ in the near future. The use of atomic internal levels may be replaced by motional levels of each of the atoms, if they are trapped in pseudo-harmonic potentials \cite{parkins95a}, since there can be many of them at our disposal. There are several sources of errors that have not been considered here and that will give limitations to our proposal, such as decoherence of hyperfine levels, laser fluctuations, non perfect atom-resolving detection, although we expect that they can be controlled to a large extent. Other effects such as retardation, imperfect atomic cooling, or fluctuations in atomic positions may have to be considered in the analysis to achieve large fidelities (see, however \cite{gonzaleztudela15a} where part of these errors were already considered).

In conclusion, we have presented several probabilistic protocols to generate heralded highly entangled atomic states that afterwards can be mapped to photonic states at will with very high fidelities. In particular, we show how to accumulate $m$ collective atomic excitations with infidelity $I_m=0$ and an exponential number of operations $R_m=p^{-m}$, being $p$ the heralding probability for adding a single excitation. We design a protocol where $p$ can be close to unity for systems with large $N,P_\oned$, which would allow to accumulate $m\sim 20-50$ excitations with systems $P_\oned\sim 10^2-10^3$ using $R_m\sim 10^4$ operations with unit fidelity [see Fig.~\ref{fig3}]. Moreover, we also present a protocol with polynomial scaling in the number of operations $R_m$ by using atomic excitation number resolved and overall low infidelity $I_m\propto \mathrm{poly}(m)/(N P_\oned)$. Though, we discus our protocols in the context of waveguide QED, the ideas can also be exported to other systems such as low mode 
cavities \cite{specht11a,thompson13a,tiecke14a,haas14a}, superconducting circuits \cite{mlynek14a} or optical fibers \cite{vetsch10a,goban12a,petersen14a,beguin14a} with suitable modification of the protocols. 

\textbf{Acknowledgements.}
The work of AGT, VP and JIC was funded by the European Union integrated project \emph{Simulators and Interfaces with Quantum Systems} (SIQS). AGT also acknowledges support from Intra-European Marie-Curie Fellowship NanoQuIS (625955). VP acknowledges the Cluster of Excellence NIM. HJK acknowledges funding by the Air Force Office of Scientific Research, Quantum Memories in Photon-Atomic-Solid State Systems (QuMPASS) Multidisciplinary University Research Initiative (MURI), by the Institute of Quantum Information and Matter, a National Science Foundation (NSF) Physics Frontier Center with support of the Moore Foundation by the  Department of Defense National Security Science and Engineering Faculty Fellows (DoD NSSEFF) program, by NSF PHY1205729 and support as a Max Planck Institute for Quantum Optics Distinguished Scholar.

\bibliography{Sci,books}
\bibliographystyle{naturemag}

\newpage
\widetext
\begin{center}
\textbf{\large Supplemental Material: Reliable multiphoton generation in waveguide QED}
\end{center}
\setcounter{equation}{0}
\setcounter{figure}{0}
\makeatletter

\renewcommand{\thefigure}{SM\arabic{figure}}
\renewcommand{\thesection}{SM\arabic{section}}  
\renewcommand{\theequation}{SM\arabic{equation}}

In this Supplementary Material, we provide the full details on how to characterize the different protocols described within the main manuscript. For the sake of understanding, we devote different Sections to the each protocol including in all of them the atom-waveguide tools required and the calculation of probabilities and fidelities such that each Section is self-contained. The summary of the Supplementary Material reads as follows:
\begin{itemize}
 \item In Section \ref{sec:DLCZ}, we discuss the first protocol of the main manuscript which adapts a protocol envisioned for entangling atomic ensembles \cite{duan01a} to herald the generation of $m$ collective excitations using single photon detections within the waveguide. We discuss both the situation where we accumulate directly the excitations into the same atomic level and where we store it in another hyperfine level and combine them a posteriori. 
 
 \item In Section \ref{sec:heralded0}, we discuss the second protocol of the main manuscript which accumulate excitations directly within the same level using a two-step protocol. The protocol overcomes the trade-off between probability of success and fidelities of the first one but still requires an exponential number of operations.
 
 \item In Section \ref{sec:heralded1}, we discuss the third protocol which show to overcome the exponential scaling of the number of operations of the second protocol by using extra auxiliary levels to store the excitations after each heralded transfer while keeping the overall infidelity still low
 
 \item For completeness, in Section \ref{sec:heralded2} we discuss a variation of the previous two protocol that relaxes part of the requirements of the protocols two and three, e.g., closed transition, at expense of slightly worse scaling of probabilities and infidelities.
 
 \item Finally, in Section \ref{sec:doubling} we discuss on how to use beam splitters transformations and atomic detection to merge atomic excitations and reach high photon numbers.  
\end{itemize}

\section{Heralding multiple collective atomic excitations in waveguide QED setups using single photon detectors. \label{sec:DLCZ}}

In this Section is show how to adapt a protocol originally envisioned to generate entanglement between distant ensembles~\cite{duan01a}, to create multiphoton states in waveguide QED setups. The idea is to accumulate several collective atomic excitations, that can afterwards be mapped to $m$-photon states with very high fidelities \cite{porras08a,gonzaleztudela15a}. 

The protocol consists in heralding the transfer a single collective atomic excitations $m$ times through $m$ photon detection events within the waveguide. We discuss two different approaches: i) one where we store the heralded single excitation in a different level $s_1$ with methods that we revise in Section \ref{sec:doubling}; ii) or where we store the excitation directly into the same level. 

The key step in both cases is to study how an additional single collective excitation can be added to an atomic ensemble which already contains $m$ excitations in another level, $s_1$, i.e.
\begin{equation}
\label{eqSM:phini2}
\ket{\Phi_m^{s_1}}=\frac{1}{\sqrt{\binom{N}{m}}} \mathrm{sym}\{\ket{s_1}^{\otimes m}\ket{g}^{\otimes (N-m)}\}\,.
\end{equation}

Notice, that due to low heralding efficiencies $p$ of these kind of protocols \cite{duan01a}, it is not clear whether the decoherence introduced when repeating the protocol $1/p$ times will spoil the coherence of the already stored excitations in $\ket{\Phi_m^{s_1}}$. In the following Sections, we analyze what are the atom-waveguide resources that we use to extend the protocol for multiple excitations. Then, we analyze the probabilities of the different processes that may occur in each attempt of generating an extra collective excitation, and finally, we take everything into account to estimate the fidelity of adding an extra excitation when the system is already in $\ket{\Phi_m^{s_1}}$. In a separate Section, we will consider the modifications of probabilities when we directly accumulate in $s$, such that the initial state is $\ket{\Phi_m^{s}}$.

\subsection{Atom-waveguide resources}

\begin{figure}[bt]
	\centering
	\includegraphics[width=0.95\textwidth]{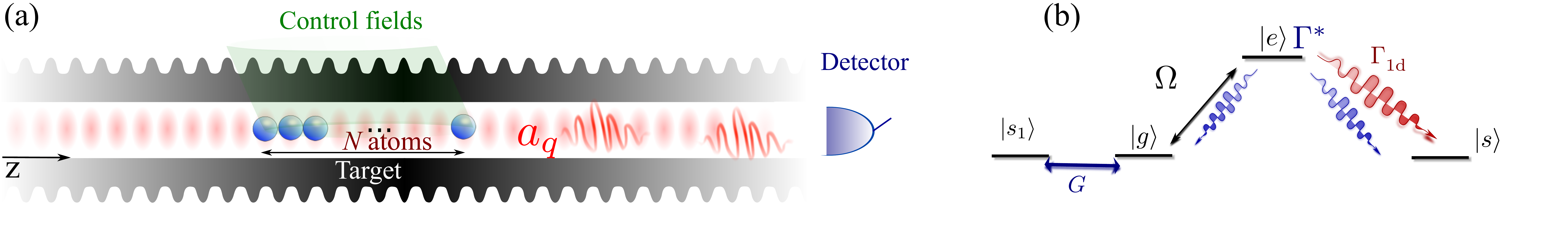}
	\caption{(a) General setup for first protocol discussed in the main manuscript which consists of a collective ensemble where we prepare superpositions and a single photon detector that we use to herald the excitations. (b) Internal level structure of emitters in which the transition $\ket{e}\leftrightarrow\ket{s}$ is coupled to a waveguide mode with rate $\Gamma_{\oned}$. The transition $\ket{g}\leftrightarrow \ket{e}$ is controlled through the Raman laser $\Omega$. We also include the possibility of having an extra auxiliary level $\ket{s_1}$ where excitations can be stored using microwave [or two-photon Raman] fields $G$.}\label{figSM1bis}
\end{figure}

The tools/resources that we use are:
\begin{itemize}
\item We assume that we have $N$ \emph{target} atoms, in which we prepare the superposition, trapped close to a one-dimensional waveguide as depicted in Fig.~\ref{figSM1bis} (a). Moreover, we assume that the atoms have an internal level structure, as shown in Fig.~\ref{figSM1bis}(b), where one optical transition $\ket{e}\leftrightarrow\ket{s}$ is coupled to a waveguide mode with a rate $\Gamma_\oned$, and the other leg $\ket{e}\rightarrow \ket{g}$ is controlled by a classical field $\Omega^n$, with $n=1,\dots,N$. In one of the schemes that we explore, we also assume that we have an extra mode $s_1$, where the excitations can be stored and which can be combined through microwave or two-photon Raman transitions with Rabi frequency $G$.

\item The atoms are placed at distances commensurate with the characteristic wavelength of the guided mode, $\lambda_\aa$, at the frequency of the transition $\ket{e}\rightarrow \ket{g}$, i.e. $z_n=n\lambda_\aa$. With that assumption, it is easy to show that effective atom dynamics induced by the interaction with the waveguide is solely driven by long-range dissipative couplings:
\begin{align}
\mathcal{L}_{\mathrm{coll}}\left[ \rho \right] = \frac{\Gamma_{\oned}}{2} D_{ S_{ge}}[\rho]\,,
\end{align}
where $\rho$ is the reduced density matrix of the atomic system, $D_O[\rho]=(2 O\rho O^\dagger -O^\dagger O\rho-\rho O^\dagger O)$ is the dissipator associated to a given jump operator $O$, and where we defined the collective spin operator of the target ensemble as $S_{\alpha\beta} = \sum_{n=1}^N \sigma_{\alpha\beta}^{n}$, with $\sigma^{n}_{\alpha\beta}=\ket{\alpha}_n\bra{\beta}$.  Obviously, the excited state can also decay to other modes other than into the waveguide, which leads to extra Lindblad terms which read:
\begin{align}
\mathcal{L}_{*}(\rho)=\sum_{n=1,\dots,N}\left(\frac{\Gamma^*}{4} D_{\sigma^n_{se} }[\rho]+\frac{\Gamma^*}{4} D_{\sigma^n_{ge} }[\rho]\right)\,,
\end{align}
which lead to finite Purcell factor $P_\oned=\frac{\Gamma_\oned}{\Gamma^*}$.

\item A single photon detector with overall efficiency $\eta$ at the end of the waveguide to detect the emission of waveguide photons.

\end{itemize}

\subsection{Protocol and calculation of probabilities.}

The protocol works as follows: we apply a short laser pulse $\Omega^n=\Omega$ such that we drive the collective dipole excitation coupled to the waveguide $S_{ge}$, during a short time $T$ such that $\sqrt{N}\Omega T/2=x\ll 1$. 
Then, the stored superposition $\ket{\Phi_m^{s_1}}$ evolves into:
\begin{align}
 \label{eqSM:dlczini}
 \ket{\Psi_0}\propto \ket{\Phi_m^{s_1}}+x\frac{1}{\sqrt{N_m}}S_{eg}\ket{\Phi_m^{s_1}}+\frac{x^2}{2}\sqrt{\frac{2}{N_m(N_m-1)}}S_{eg}^2\ket{\Phi_m^{s_1}}+O(x^3)\,.
\end{align}

If we leave the system free to decay for a time $t\gg (\Gamma^*)^{-1}$, the system evolves to (dropping normalization and coefficients):
\begin{align}
 \label{eqSM:dlczini2}
 \ket{\Psi_0}\rightarrow & \ket{\Phi_m^{s_1}}+x S_{sg} \ket{\Phi_m^{s_1}}\otimes \ket{1_k}+x\sum_n \sigma_{gg}^n \ket{\Phi_m^{s_1}}\otimes \ket{1^*_{k,n}} +\sum_n \sigma_{sg}^n \ket{\Phi_m^{s_1}}\otimes \ket{1^*_{k,n}} +\,,\nonumber \\
 &+x^2S_{sg}^2 \ket{\Phi_m^{s_1}}\otimes \ket{2_k}+x^2\sum_n \sigma_{gg}^n S_{sg} \ket{\Phi_m^{s_1}}\otimes \ket{1_k}\otimes \ket{1^*_{k,n}}+x^2\sum_n \sigma_{sg}^n S_{sg} \ket{\Phi_m^{s_1}}\otimes \ket{1_k}\otimes \ket{1^*_{k,n}}+\nonumber \\
 &+x^2\sum_{n,m} \sigma_{sg}^n \sigma_{gg}^m \ket{\Phi_m^{s_1}}\otimes  \ket{1^*_{k,m}}\otimes \ket{1^*_{k,n}}+x^2 \sum_{n,m} \sigma_{gg}^n \sigma_{gg}^m \ket{\Phi_m^{s_1}}\otimes  \ket{1^*_{k,m}}\otimes \ket{1^*_{k,n}}+x^2\sum_{n,m} \sigma_{sg}^n \sigma_{sg}^m \ket{\Phi_m^{s_1}}\otimes  \ket{1^*_{k,m}}\otimes \ket{1^*_{k,n}}\,,
\end{align}
where the $\ket{M_k}$ [$\ket{M_k^*}$] denotes waveguide (leaky) photon states. The first line of the equation contains the processes where no photon ($ \ket{\Phi_m^{s_1}}$) or only one photon is emitted either through the waveguide or through spontaneous emission. In the second and third lines we write the different processes associated with the existence of double quantum jumps. In order to obtain the different probabilities associated to each process, we use the expansion of the master equation \cite{gardiner_book00a} distinguishing the evolution of the effective non-hermitian Hamiltonian, i.e., $S(t,t_0)\rho=e^{-i H_{\mathrm{eff}} t}\rho e^{i H_{\mathrm{eff}}^\dagger t}$, and the one resulting from quantum jumps evolution, i.e., $J\rho$, that is given in general by:
\begin{align}
 \rho(t)= S(t,t_0)\rho(t_0)+  \sum_{n=1}^\infty \int_{t_0}^t dt_n S(t,t_n)J(t_n)\dots  \int_{t_0}^{t_2} dt_1S(t_1,t_0)\rho(t_0) \,.
\end{align}

The non-hermitian evolution leads to:
\begin{align}
 \label{eqSM:dlczininh}
 \ket{\Psi(t)}\propto e^{-i H_{\mathrm{eff}} t}\ket{\Psi_0}\propto \ \ket{\Phi_m^{s_1}}+x\frac{e^{-(\Gamma_\oned+\Gamma^*)t}}{\sqrt{N_m}}S_{eg}\ \ket{\Phi_m^{s_1}}+\frac{x^2}{2}e^{-2(\Gamma_\oned+\Gamma^*)t}\sqrt{\frac{2}{N_m(N_m-1)}}S_{eg}^2 \ket{\Phi_m^{s_1}}+O(x^3)\,,
\end{align}
where we see that if we wait a time long enough, i.e., $t\gg (\Gamma^*)^{-1}$ the only population remaining in $\ket{\Psi(T)}\rightarrow \ \ket{\Phi_m^{s_1}}$. The probability emitting a collective photon from $S_{eg} \ket{\Phi_m^{s_1}}$ is given by:
\begin{equation}
 \label{eq:colljump}
 p_{\mathrm{coll}}=\frac{x^2}{1+\frac{1}{P_\oned}}\approx x^2\left(1-\frac{1}{P_\oned}\right)\,,
\end{equation}
where in the last approximation we assumed to be in a regime with $P_\oned\gg 1$, whereas the one of emitting an spontaneous emission photon from the same state:
\begin{equation}
 \label{eq:jump}
 p_{*}=\frac{x^2}{P_\oned+1}\approx \frac{x^2}{P_\oned}\,.
\end{equation}

We herald with the detection of a waveguide photon with efficiency $\eta$, such that the final success probability of heralding is:
\begin{equation}
p= p_{\mathrm{coll}}\times \eta\approx  \eta x^2\left(1-\frac{1}{P_\oned}\right)\,,
\end{equation}

In case of no detection it may have happened that our waveguide photon was emitted but not detected, that is, the atoms will indeed be in a superposition $S_{sg}\ \ket{\Phi_m^{s_1}}$. So, before making a new attempt, we need to ensure that we pump back any possible excitation in $\ket{g}$ to $\ket{s}$.  However, to minimize the probability of emitting an leaky photon, which spoils the coherence of our stored superpositions, one needs to apply a repumping procedure in three steps:
\begin{itemize}
\item First, apply a $\pi$ pulse with a microwave field that flips all the excitations $\ket{g}\rightarrow \ket{s}$. This switches $S_{sg}\ket{\Phi_m^{s_1}}$ into $S_{gs} \ket{\Psi_m^{s_1}}$, where $\ket{\Psi_m^{s_1}}$ is the same as $\ket{\Phi_m^{s_1}}$ but with levels $\ket{s}$ and $\ket{g}$ interchanged.

\item Then, we apply a fast Raman $\pi$-pulse with $\Omega\gg N \Gamma_\oned$ such that $S_{gs} \ket{\Psi_m^{s_1}}\rightarrow S_{es}\ket{\Psi_m^{s_1}}\rightarrow \ket{\Psi_m^{s_1}}$. This incoherent transfer is done through a collective photon, such that the probability of emitting a leaky photon is:
\begin{equation}
\label{eqs:incpump}
p_{\mathrm{pump},*}\approx (1-\eta)\times p_{\mathrm{coll}} \times \frac{1}{N_m P_\oned}\approx (1-\eta)\frac{x^2}{N_m P_\oned}\,.
\end{equation}

\item Finally, we reverse the microwave $\pi$-pulse such that $\ket{g}\leftrightarrow \ket{s}$ and therefore $ \ket{\Psi_m^{s_1}}\rightarrow  \ket{\Phi_m^{s_1}}$.

\end{itemize}

\subsection{Fidelities of the protocol.}

In order to calculate the errors (and fidelities) of the protocol we have to distinguish between the errors introduced after successfully heralding the excitations, and the error per trial that we introduce when we repeat the protocol after failure. 
\begin{itemize}
 \item \emph{Successful heralding:} From Eq.~\ref{eqSM:dlczini2}, it can be seen that if we detect a single photon in the waveguide, this means that the initial density matrix $\rho_m=\ketbra{\Psi_m}$ transforms into:
 \begin{align}
 \label{eqSM:DLCZ2}
  \rho_m\rightarrow \frac{1}{N_m}S_{gs}\rho_m S_{sg}+\frac{2 x^2 (1-\eta_\rd)}{N_m(N_m-1)}S_{gs}^2\rho_m S_{sg}^2\,,
 \end{align}
 where the first term is the desired process, whereas the second corresponds to the probability of detecting only one of the two photons emitted from the doubly excited terms $S_{es}^2\ \ket{\Phi_m^{s_1}}$. This introduces a large error as the state $S_{gs}\ \ket{\Phi_m^{s_1}}$ is orthogonal to the state that we want to create, such that the error when heralding is:
\begin{equation}
\label{eq:errdouble}
\varepsilon_{\mathrm{double}}=x^2(1-\eta)\,.
\end{equation}

\item  \emph{Failed heralding:} If we detect no photon, then, our state is projected to:
 \begin{align}
  \rho_m\rightarrow (1-x^2) \rho_m +p_{\mathrm{coll}}(1-\eta)\frac{1}{N_m}S_{gs}\rho_m S_{sg}+p_{*} \mathbb{J}_{*}\rho_m+O(x^4)\,,
 \end{align}
which corresponds to processes in which: i) we have not created any excitation in the system, with probability ($1-x^2$), ii) we have created a single collective excitation emitting a collective photon but we have not detected it; iii) we have created a single excitation, but it has emitted a free space photon, represented through $\mathbb{J}_{*}\rho_m$. As we explained before, after appropriate repumping, the errors from undetected collective quantum jumps can be corrected introducing some extra spontaneous emission probability, such the final density matrix after repumping is given by:
 \begin{align}
  \rho_m\rightarrow (1-x^2) \rho_m +\left(p_{\mathrm{pump},*}+p_{*}\right) \mathbb{J}_{*}\rho_m+O(x^4)\,,
 \end{align}

 Fortunately, the errors from spontaneous emission are not so severe as the resulting state still have a big overlap with the original state:
 \begin{align}
\bra{\Phi_m}\mathbb{J}_{*}[\rho_m]\ket{\Phi_m}=\frac{\binom{N-1}{m}}{\binom{N}{m}}\approx 1-\frac{2 m }{N}\,.
\end{align}

Therefore, the error introduced per failed attempt is given by:
\begin{equation}
\label{erfailDLCZ}
\varepsilon_{\mathrm{fail},*}=\left(p_{\mathrm{pump},*}+p_{\*}\right)\frac{ m }{N}
\end{equation}

\item \emph{Complete process: } The complete process consists (in average) of a successful heralding event and $1/p$ repetitions such that the average final (in)fidelity to generate $S_{sg} \ket{\Phi_m^{s_1}}$ is given by:
  \begin{equation}
   \label{eqDLCZ:error}
   I_{m\rightarrow m+1}\approx \frac{\varepsilon_{\mathrm{fail},*}}{p}+\varepsilon_{\mathrm{double}}\approx \frac{m}{\eta N P_\oned}+(1-\eta) x^2\,,
  \end{equation}

\end{itemize}

Therefore, the best fidelity that can be obtained is done by imposing: $x^2=\frac{m}{\eta(1-\eta) N P_\oned}$, however, at the price of bad scaling of probability: $p=\frac{m}{N P_\oned (1-\eta)}$ \footnote{All the expressions are valid obviously for detectors with $0<\eta<1$}. This trade off between probabilities and fidelities comes from the well known problem of double excitations in these type of probabilistic protocols for atomic ensembles. The total error to accumulate $m$ excitations depends on how to combine excitations from $s_1$ and $s$ and will be discussed in Section \ref{sec:doubling}.

\subsubsection{Accumulating excitations in the same level.}

If we accumulate excitations directly in $s$, the initial state in each step will be $\ket{\Phi_m^{s}}$ instead of $\ket{\Phi_m^{s_1}}$. The protocol works in the very same way, but the heralding probability has in this case a small correction in $m$ which reads:
\begin{equation}
p_m= p_{\mathrm{coll},m}\times \eta\approx  \eta x^2\left(1-\frac{1}{(m+1) P_\oned}\right)\,.
\end{equation}

Moreover, in case of failure we reinitialize the process from the beginning, pumping back all the atoms to $g$, which avoids the need of applying the repumping protocol discussed below. The infidelity at each step in this case only come from double excitations contributions scale as:
  \begin{equation}
   \label{eqDLCZ:error2}
   I_{m\rightarrow m+1}= \varepsilon_{\mathrm{double}}\approx (1-\eta) x^2\,.
  \end{equation}

The total error to accumulate $m$ collective excitations will be directly $I_m=m(1-\eta) x^2$. 

\section{Heralding single collective excitations using atom-waveguide QED: two-step protocol with exponential scaling. \label{sec:heralded0}}

In this Section we discuss the second protocol of the main manuscript which overcomes the trade-off between heralding probabilities and fidelities. The starting point is that our target ensemble already contains $m$ collective excitation in a given level $s$, i.e., our initial state is $\ket{\Phi_m^{s}}$. Then, the protocol is divided in two-steps: i) we use a single atom to provide single excitations to the target ensemble; ii) then, we use an independent ensemble to herald the successful transfer of the excitations to the target ensemble.

\subsection{Atom-waveguide resources.}

\begin{figure}[bt]
	\centering
	\includegraphics[width=0.95\textwidth]{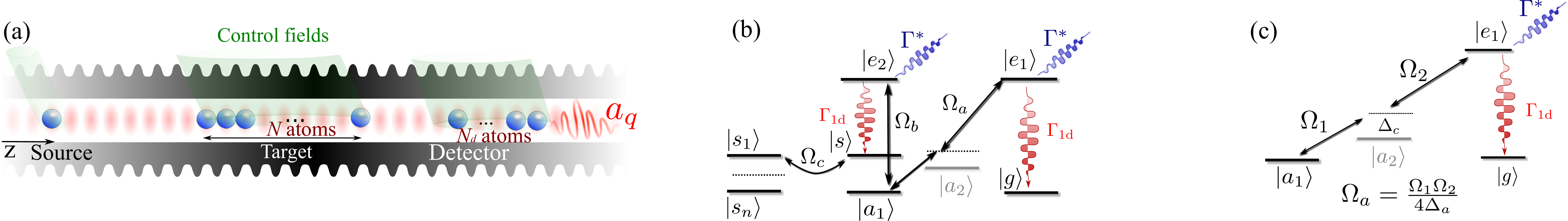}
	\caption{(a) General setup consisting in three individually ensembles: a single \emph{source} atom and a \emph{target}/\emph{detector} ensembles and $N/N_d$ atoms respectively. (b) Internal level structure of emitters in which the transition $e_1 (,e_2)\leftrightarrow g (,s)$ is coupled to a waveguide mode with rate $\Gamma_\oned$. The transition $e_1\leftrightarrow a_1$ is driven by a two-photon off resonant process such that there is no spontaneous emission associated to it. The transition $e_2\rightarrow s$ is controlled through a laser with Rabi frequency $\Omega_b$, and the transition $s\leftrightarrow s_n$ is controlled through microwave/Raman laser $\Omega_c$. In the case of the target atoms, we require extra hyperfine levels $\{\ket{s_m}\}_n$, to store atomic excitations. (c) Detail of closed transition implementation using a two photon Raman transition using $\ket{a_2}$ as the auxiliary intermediate state.}\label{figSM1}
\end{figure}

The requirements that we need for this protocol are:
\begin{itemize}

\item The general setup is depicted in Fig.~\ref{figSM1}(a): we have one individually addressable \emph{source} atom, and two independently addressable ensembles with $N,N_d$ atoms respectively, one where we store the collective excitations, namely, the $\emph{target}$ ensemble and another one that we use to herald the excitations, namely, the $\emph{detector}$ ensemble. We assume that the atoms have a general level structure as depicted in Fig.~\ref{figSM1}(b), with two dipolar transitions $e_1 (,e_2)\leftrightarrow g (,s)$ coupled to a waveguide modes with an effective rate $\Gamma_{\oned}$.

\item The atoms must be placed at distances $z_n=n \lambda_\aa$, with $n\in \mathbb{N}$ and being $\lambda_\aa$ the characteristic wavelength of the guided mode at the atomic frequency. With that choice, the dynamics induced by the waveguide is given only by long-range dissipative couplings, described by the following Lindblad terms:
\begin{align}
\mathcal{L}_{\mathrm{coll}}\left[ \rho \right] = \frac{\Gamma_{\oned}}{2} D_{ \sigma_{ge_1}^\rs+ S_{ge_1}+S_{ge_1}^d}[\rho]+\frac{\Gamma_{\oned}}{2} D_{ \sigma_{se_2}^\rs+S_{se_2}+S_{se_2}^d}[\rho]\,.
\end{align}
where $S_{\alpha\beta},S^d_{\alpha\beta}$ denote the collective spin operators ($\sum_n \sigma^n_{\alpha\beta}$) in the target/detector atoms. Note that in principle they will be expressed all in a collective jump operator as they are all coupled to the same guided mode. However, we separate them for convenience as in our protocol each term will act in the two different steps of the protocol.

\item The excited states may also decay to other channels than the one inducing the long-range couplings, which we embedded in a single decay rate $\Gamma^*$. However, to obtain the desired scaling we use a closed transition \cite{vanenk97a,porras08a,borregaard15a}, that we study more in detail in a separate section. Thus, the final spontaneous emitted photons are described by:
\begin{equation}
\label{eqSM2:mequation2}
{\cal L}_{*}(\rho) = 
\sum_{n}\Big(\frac{\Gamma^*}{4} D_{\sigma^n_{se_2} }[\rho]+\frac{\Gamma^*}{4} D_{\sigma^n_{a_1 e_2} }[\rho]+\frac{\Gamma^*}{2} D_{\sigma^n_{g e_1} }[\rho]\Big)\,,
\end{equation}
where we define the coefficients accompanying the jump operators, such that both transitions have the same $P_{\oned}$.

\item Finally, we also require the existence of Raman/microwave fields, $\Omega_{a,b,c}$ as depicted in Fig.~\ref{figSM1}(a) that classically control the transition between different levels.
\end{itemize}

\subsubsection{Example of closed transition}

There have already been several proposals in the literature of closed transitions \cite{vanenk97a,porras08a,borregaard15a} but for completeness, we discuss here in detail one of particular realization sketched in Fig.~\ref{figSM1}(c). The state $a_1$ is connected to the excited state $e_1$ with a two-photon transition through the intermediate state $a_2$. The level $a_1$ is chosen, e.g., such that the direct transition $a_1\leftrightarrow e_1$ is dipole forbidden. Then, denoting as $\Omega_1(\Omega_2)$ the Rabi frequency connecting $a_1\leftrightarrow a_2 (,a_2\leftrightarrow e_1)$, and choosing a detuning $\Delta_a\gg \Omega_1,\Omega_2$, the microwave and Raman laser induce a two photon transition from $a_1\leftrightarrow e_1$ with effective driving $\Omega_{a}=\frac{\Omega_1\Omega_2}{4\Delta_a}$.

\subsection{Protocol and calculation of probabilities}

The protocol for heralding single collective excitation starts with the source/target/detector atoms in $\ket{a_1}_s\otimes \ket{\Phi_m^{s}}_t \otimes \ket{s}^N_d$ and it consists of several steps (as depicted in Fig.~2b of the main manuscript):
\begin{enumerate}

\item In the first step, we move the target/detector atoms in $s$ to an auxiliary level $s_1$, such that we have $\ket{a_1}_s\otimes \ket{\Phi_m^{s_1}}_t \otimes \ket{s_1}^{N_d}$, avoiding that the excitations from the waveguide can be absorbed the detector level $s$.
 \item In the second step, we excite the source atom to $e_1$, then, we switch on $\Omega_a^t$ under the Quantum Zeno dynamics conditions ($\Omega_b^d\ll N_d P_\oned$), such that the excitation is coherently transferred to the ensemble to $a_1$, i.e., $S_{a_1 g}\ket{\Phi_m^{s_1}}$.
 \item Now, we move back the states in the target/detector ensemble from $s_1$ to $s$, such that, we prepare $S_{a_1 g}\ket{\Phi_m^{s}}_t \otimes \ket{s}^{\otimes N_d}_d$. Moreover, we move the source atom to, e.g., $s_1$, such that it plays no role in subsequent steps.
 \item Then, we do a fast $\pi$-pulse in the ensemble with $\Omega_b$, such that, $S_{a_1 g}\ket{\Phi_m^{s}}\rightarrow S_{e_2 g}\ket{\Phi_m^{s}}$. Afterwards, we apply $\Omega_b^d$ within the Zeno dynamics conditions ($\Omega_b^d\ll N_d P_\oned$), such that we arrive to a state in the target/detector atoms $\ket{\Phi_{m+1}^{s}}\otimes S_{a_1 s}\ket{s}^{\otimes N_d}$.
 
 \item Thus, if we measure the detector atoms in $a_1$, we herald the transfer of a single collective excitation to $s$.
 
\end{enumerate}

Assuming that $\pi$-pulses are perfect, the relevant steps of the protocols for the analysis of probabilities and fidelities are the second and the fourth. Interestingly, they can be analyzed in a common way as in each of them what happens is that a single excitation is transferred through the waveguide via Zeno dynamics to an ensemble with $N/N_d$ atoms respectively. Therefore, in the following Section we analyze the general problem and then we will particularize for steps 2 and 4.

\subsubsection{General Zeno step}

The general problem consists of two ensembles (a and b) with three-level atoms (with metastable states $\ket{0}$, $\ket{1}$ and excited state $\ket{2}$) that contain $N_\mathrm{a}$ and $N_\mathrm{b}$ atoms each. The dynamics are governed by the collective decay on the $\ket{2}-\ket{1}$ transition, i.e., $\mathcal{L} \left[\rho\right] = \frac{1}{2} \Gamma_\oned D_{S_{12}^{\mathrm{(a)}} + S_{12}^{\mathrm{(b)}}}[\rho]$ and an external field on the second ensemble, i.e., $H_\mathrm{b} = \frac{1}{2} \Omega S_{21}^{\mathrm{(b)}} + \mathrm{h.c.}$. Therefore the effective non-hermitian hamiltonian that drives the no-jump evolution is given by:
\begin{equation}
H_\mathrm{eff} = 
	\frac{1}{2} \left(\Omega S_{21}^{\mathrm{(b)}} + \mathrm{h.c.} \right)
	-\mathrm{i} \frac{\Gamma_\oned}{2}
		\big(S_{21}^{\mathrm{(a)}} + S_{21}^{\mathrm{(b)}} \big) 
		\big(S_{12}^{\mathrm{(a)}} + S_{12}^{\mathrm{(b)}} \big)
	-\mathrm{i} \frac{\Gamma^*}{2}\sum_{n}\sigma^{n}_{ee}.
\end{equation}

Denoting the collective symmetric excitations as $\ket{\#_0, \#_1, \#_2}_\mathrm{a/b}$, the initial state of the problem we want to analyze can be written $\ket{0,0,1}_\mathrm{a} \otimes \ket{0,N-m,0}_\mathrm{b}$ and the goal state is $\ket{0,1,0}_\mathrm{a} \otimes \ket{1,N-m-1,0}_\mathrm{b}$.

The decay operators couple an initial state $\ket{\psi_1} = \ket{N_\aa - k - 1, k, 1}_\aa \otimes \ket{0, N_b, 0}$ to $\ket{\psi_2} = \ket{N_\aa - k - 1, k+1, 0}_\aa \otimes \ket{0, N_b, 1}$ and the coherent terms couple the latter to $\ket{\psi_3} = \ket{N_\aa - k - 1, k+1, 0}_\aa \otimes \ket{1, N_b, 0}$. In this basis, one can write the Hamiltonian as
\begin{equation}
H_\mathrm{eff} = 
\frac{1}{2} \left( \begin{array}{ccc} 
	-i \left((k+1)\Gamma_\oned+\Gamma^*\right) & -i \sqrt{(k+1) N_\bb} \Gamma_\oned & 0 \\
	-i \sqrt{(k+1) N_\bb} \Gamma_\oned  & - i (N_\bb \Gamma_\oned+\Gamma^*) & \Omega \\
	0 & \Omega & 0
\end{array} \right).
\end{equation}
or equivalently in the basis of superradiant and dark states, that is $\{ \ket{\psi_s} , \ket{\psi_d}, \ket{\psi_3}\}$ with $\ket{\psi_s}  = \sqrt{\frac{k+1}{N_\bb + k+1}} \ket{\psi_1} + \sqrt{\frac{N_\bb}{N_\bb + k+1}} \ket{\psi_2} $ and $\ket{\psi_d} = \sqrt{\frac{N_b}{N_\bb + k+1}} \ket{\psi_1} - \sqrt{\frac{m+1}{N_\bb + k+1}} \ket{\psi_2}$, as
\begin{equation}
\widetilde{H}_\mathrm{eff} = 
\frac{1}{2} \left( \begin{array}{ccc} 
	-i \left((N_\bb + k + 1) \Gamma_\oned + \Gamma^*\right) & 0 & -\sqrt{\frac{N_\bb}{N_\bb + k+1}} \Omega \\
	0 & -i \Gamma^* & -\sqrt{\frac{k+1}{N_\bb + k+1}} \Omega   \\
	-\sqrt{\frac{N_\bb}{N_\bb + k+1}} \Omega & -\sqrt{\frac{k+1}{N_\bb + k+1}} \Omega   & 0
\end{array} \right)\,.
\end{equation}

When the coherent driving is weak compared to the collective dissipation $\Omega \ll N_\bb \Gamma_\oned$, the superradiant state can be adiabatically eliminated and an effective Zeno Dynamics between the dark states is obtained, which is governed by the Hamiltonian
\begin{equation}
H_\mathrm{AE} \approx
	\frac{1}{2} \left( \begin{array}{ccc} 
		-i \left((N_\bb + k + 1) \Gamma_\oned + \Gamma^*\right) & 0 & 0 \\ 
		0 & -i \Gamma^* &  -i \sqrt{\frac{k+1}{N_\bb + k+1}} \Omega  \\
		0 & -i \sqrt{\frac{m+1}{N_\bb + k+1}} \Omega  & -i \frac{N_\bb |\Omega|^2}{(N_\bb + k + 1)^2\Gamma_\oned} 
	\end{array} \right).
\end{equation}

To minimize the errors, one chooses $\Omega = \sqrt{(N_\bb + k +1) \Gamma_\oned \Gamma^*}$ such that $\frac{N_\bb |\Omega|^2}{(N_\bb + k + 1)^2\Gamma_\oned} = \Gamma^*$. The dynamics for each state is then approximately given by
\begin{align}
&|\psi_d(t)|^2\approx 
	\frac{N_\bb}{N_\bb + k+1}e^{-\Gamma^* t}\cos^2\Big(t \frac{\Gamma^*\sqrt{(k+1)P_\oned}}{2}\Big),\\
&|\psi_3(t)|^2\approx 
	\frac{N_\bb}{N_\bb + k+1}e^{-\Gamma^*t }\sin^2\Big(t \frac{\Gamma^*\sqrt{(k+1)P_\oned}}{2}\Big)\equiv p,\\
&|\psi_s(t)|^2\approx \frac{k+1}{N_\bb + k+1}e^{-(\Gamma^*+(N_\bb + k + 1) \Gamma_{\oned})t}.
\end{align}

The success probability $p=|\psi_3(T)|^2$ is the maximized for $T = \pi \left( \sqrt{\frac{k+1}{N_\bb + k+1}} \Omega \right)^{-1}$ and $\Omega = \sqrt{(N_\bb + k +1) \Gamma_\oned \Gamma^*}$ for $k\ll N_\bb$. In this case, one obtains
\begin{equation}
	p = \frac{N_\bb}{N_\bb + k+1} \ee^{- \pi / \sqrt{(k+1) P_\oned}},
\end{equation}
where the prefactor originates in the non-unit overlap of the initial state with the dark state, i.e., $|\langle \psi_d |\psi_1\rangle |^2 = \frac{N_\bb}{N_\bb + k+1}$.

For the repumping protocols that we will study in Section \ref{sec:heralded1} it is important to know the probability of spontaneous jumps in both ensembles during the evolution. As we already saw in the first protocol, the problematic processes are the one associated to leaky photons. The quantum jump analysis shows that the probability for a spontaneous jump in the ensemble $a$ or $b$ is given by
\begin{align}
p_{\aa,*}=p_{\aa_1,*}+p_{\aa_2,*}=\Gamma^*\int_0^{T} \rd t_1 |\psi_1(t_1)|^2
	+ \Gamma^*\int_{0}^\infty \rd t_1 |\psi_1(T)|^2 e^{-\Gamma^* t_1}, \\
p_{\bb,*}=p_{\bb_1,*}+p_{\bb_2,*}=\Gamma^*\int_0^{T} \rd t_1 |\psi_2(t_1)|^2
	+ \Gamma^*\int_{0}^\infty \rd t_1|\psi_2(T)|^2 e^{-\Gamma^* t_1},
\end{align}
where the first parts, $p_{\aa,\bb_1,*}$, corresponds to the interval of time $(0,T)$ where $\Omega$ is switched on, and the second part, $p_{\aa_2,*}$, the time $t\gg 1/\Gamma^*$ that we wait, with  $\Omega=0$, such that all the population in the excited state, if any, disappears. By using the approximations for $H_\mathrm{eff}$, we can calculate the different contributions and upper bound the probabilities from these processes
\begin{align}
p_{\aa_1,*}
	\lesssim \frac{1}{2}(1-e^{-\pi / \sqrt{(k+1) P_\oned}})
	\lesssim \frac{\pi}{2 \sqrt{P_\oned}}\,\\
p_{\bb_1,*}
	\lesssim \frac{k+1}{2 N_\bb}(1-e^{-\pi / \sqrt{(k+1) P_\oned}})
	\lesssim \frac{\pi \sqrt{k+1}}{2 N_\bb\sqrt{P_\oned}}\,,
\end{align}
which mainly comes from the contribution of the dark state and where the last approximation is valid for $P_\oned\gg 1$.

Finally, we need to consider what happens with the contribution $p_{\aa_2,*}$ when $P_\oned\gg 1$, and when we assume a perfect timing,  $T=\pi/\sqrt{(k+1) \Gamma_\oned \Gamma^*}$ such that then $|\psi_d(T)|^2 = 0$. The only contribution remaining is the one of the superradiant 
$\ket{\psi_s(T)} = \sqrt{\frac{k+1}{N_\bb + k+1}} e^{-(\Gamma^*+(N_\bb + k + 1) \Gamma_{\oned})T/2}$, which leads to
\begin{align}
p_{\aa_2,*}\lesssim \frac{(k+1)^2}{(N_\bb + k+1)^2} e^{-\pi (N_\bb + k + 1) / \sqrt{(k+1)P_{\oned}}}\,, \\
p_{\bb_2,*}\lesssim \frac{N_\bb (k+1)}{(N_\bb + k+1)^2} e^{-\pi (N_\bb + k + 1) / \sqrt{(k+1)P_{\oned}}} \,,
\end{align}
which are negligible compared to $p_{\aa,\bb_1,*}$ for sufficiently large $N_\bb$.

\subsubsection{Particularizing for the two step protocol: success probability and infidelity.}

The second step of the protocol that we want to analyze corresponds to identifying in the general problem $\ket{0} = \ket{a_1}$, $\ket{1}=\ket{g}$, $\ket{2}=\ket{e_1}$,  $\Omega=\Omega_\mathrm{a}$, where the ensembles have a size of $N_\mathrm{a} = 1$ and an effective size $N_\mathrm{b} = N-m$ because the $m$ excitations in all other states are decoupled and effectively reduce the atom number.  This results in a heralding probability:
\begin{align}
	p_\aa =& \frac{N-m}{N-m+1} \ee^{- \pi / \sqrt{P_\oned}}\,.
\end{align}

In the fourth step, the equivalence reads $\ket{0} = \ket{a_1}$, $\ket{1}=\ket{s}$, $\ket{2}=\ket{e_2}$,  $\Omega=\Omega_\mathrm{b}$ and the ensembles have a size of $N_\mathrm{a} = N$ and $N_\mathrm{b} = N_\mathrm{d}$. The initial state is $\ket{N-m-1,m,1}_\mathrm{a} \otimes \ket{0,N_\mathrm{d},0}_\mathrm{b}$ and the goal state is $\ket{N-m-1,m+1,0}_\mathrm{a} \otimes \ket{1,N_\mathrm{d}-1,0}_\mathrm{b}$. Thus, the success probability reads:
\begin{align}
	p_\bb =& \frac{N_\rd}{N_\rd + m+1} \ee^{- \pi / \sqrt{(m+1) P_\oned}}\,,
\end{align}
where it can be shown that in the limit of $N,N_d\gg m$, the total probability of success is lower bounded by: $p \gtrsim p_\aa^2 \approx e^{-2\pi/\sqrt{P_\oned}}$.

\subsubsection{Improving probability through pulse-shaping.}

\begin{figure}[tb]
	\centering
	\includegraphics[width=0.75\textwidth]{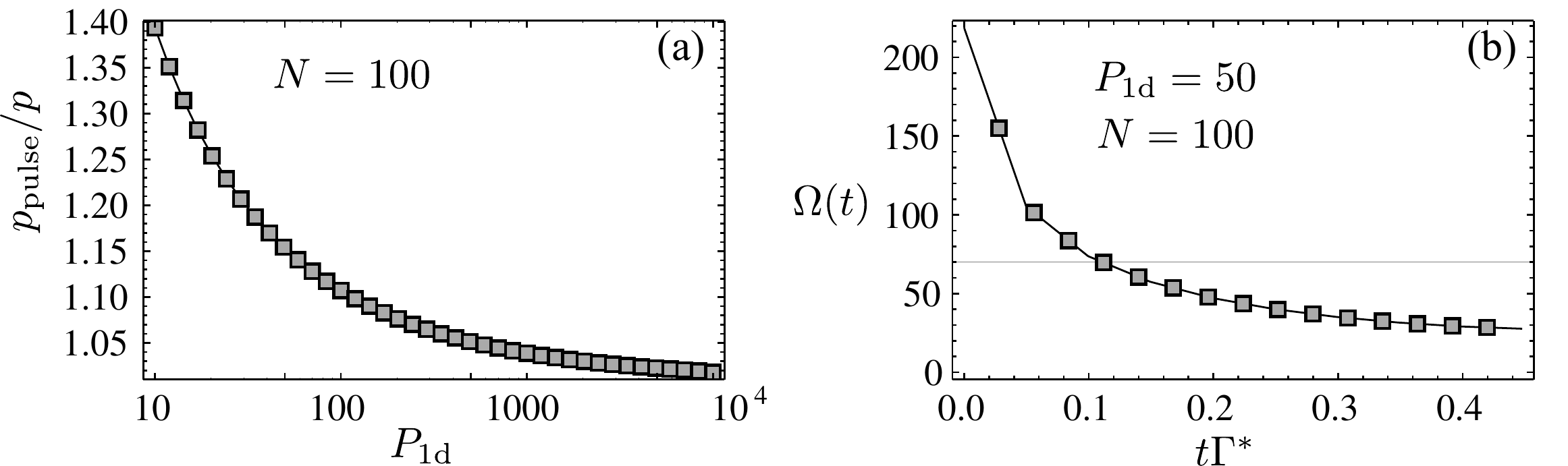}
	\caption{(a) Ratio between the probability of success with/without pulse shaping $p_\mathrm{pulse}/p$ as a function of $P_\oned$ for a situation with $N=100$ atoms in the ensemble receiving the excitations. (b) Optimal $\Omega(t)$ for a situation with $N=100$,$P_\oned=50$ using $10$ discretization steps. In red we plot the optimal $\Omega$ for the situation of constant pulses.}
	\label{figSM:pulse}
\end{figure}

So far, we have restricted our attention to the case of constant pulse driving $\Omega$. However, it is in principle possible to shape $\Omega(t)$ such that the excitation transfer from $\ket{\psi_1}$ to $\ket{\psi_3}$ is improved. The problem of optimizing $\Omega(t)$ can not be tackled analytically, however, we can choose a discretization of time $[t_1,t_2,\dots t_n]$, and try to find the optimal choice of $\Omega_i$'s such that it maximizes $|\braket{\psi_1}{\psi_3}|^2$. In Fig.~\ref{figSM:pulse}, we show an example that such optimization is possible by discretizing the optimal time found in the previous Section $T$ in $10$ steps. In Fig.~\ref{figSM:pulse}(a), we show the ratio between the probability of success with/without pulse shaping $p_\mathrm{pulse}/p$ as a function of $P_\oned$ for a situation with $N=100$ atoms in the ensemble receiving the excitations, where we show how probability can be increased, specially for moderate $P_\oned$. In panel (b), we show an example of such optimization for the case of $P_\oned=50$, showing how the optimal shape found is bigger that the optimal $\Omega_\mathrm{opt}$ (in red) for a constant pulse at initial times and then smaller. The intuition for this shape can be understood by looking the effective Hamiltonian that drives the evolution. Within the Zeno dynamics, only the dark states $\ket{\psi_d}$ and $\ket{\psi_3}$ are relevant. If we project the dynamics within this subspace, we obtain a $2\times 2$ hamiltonian:
\begin{equation}
\widetilde{H}_\mathrm{eff} \approx 
\frac{1}{2} \left( \begin{array}{cc} 
	-i \Gamma^* & \frac{\Omega}{\sqrt{N}}\\
	 \frac{\Omega}{\sqrt{N}} & -i\frac{|\Omega|^2}{N \Gamma_\oned} \\
\end{array} \right)\,.
\end{equation}
From this effective Hamiltonian, we realize that while the initial state (mostly $\ket{\psi_d}$), which only decays with $\Gamma^*$, independent of $\Omega$ the final state where we want to arrive decays with an effective decay rate which increases with $\Omega$, i.e., $\frac{|\Omega|^2}{N \Gamma_\oned}$. Therefore, it is more optimal to ramp up first $\Omega$ while the population in $\ket{\psi_3}$ is still small. By using more elaborate pulse shaping, even better improvements might be found.

\section{Heralding single collective excitations using atom-waveguide QED: two-step protocol for polynomial scaling. \label{sec:heralded1}}

In this Section we analyze the third protocol discussed in the manuscript. This protocol is a variation of the second protocol in which after each successful heralding we store the excitations in different levels $\{\ket{s_n}\}_n$ to combine them a posteriori.
This however implies the modification of the previous protocol to obtain the desired $1/(N P_\oned)$ scaling for the infidelity $I_{m\rightarrow m+1}$. In particular, the second step can not be done through Quantum Zeno dynamics, because the probability of emitting an spontaneous photon within the target ensemble is large, i.e., $p_{b,*}\propto 1/\sqrt{P_\oned}$, that would yield an infidelity $I_{m\rightarrow m+1}\propto 1/(N\sqrt{P_\oned})$. Therefore, in order to achieve subexponential scaling in the number of operations \emph{and} infidelities still scaling with $I_m\propto 1/(N P_\oned)$ one needs to make judicious modification of the setup as we will explain now.

\subsection{Atom-waveguide resources for two-step protocol.}

The requirements of this protocol are exactly analogue to the ones of Section \ref{sec:heralded0} with two modifications: i) we only require a single detector atom, i.e., $N_d=1$\footnote{Actually the source atom can also play the role of the detector atom with appropriate modification of the protocol.}; ii) we require the existence of several hyperfine states $\{s_n\}$ to store the superpositions.

\subsection{Protocol and calculation of probabilities}

In this Section, we analyze the main protocol described for heralding the transfer of a single collective excitation to the target ensemble, assuming that is already in an entangled state $\ket{\Phi_m^{\{s_n\}}}$, with $m$ excitations, which can be written as follows:
\begin{equation}
\label{eqSM:phini}
\ket{\Phi_m^{\{s_n\}}}=\frac{1}{\sqrt{\binom{N}{m_1, m_2, \dots,m_n}}} \mathrm{sym}\{\ket{s_1}^{\otimes m_1}\dots\ket{s_n}^{\otimes m_n}\ket{g}^{\otimes N_m}\}\,
\end{equation}
where $\binom{N}{m_1, m_2, \dots,m_n}=\frac{N!}{m_1!\dots m_N N_m)!}$ is the number of states within the superposition, denoting $N_m=N-m$. The final goal state we want to create is:
\begin{equation}
\label{eqSM:phigoal}
\ket{\Phi_{\mathrm{goal}}}=\frac{1}{\sqrt{N_m}} S_{sg}\ket{\Phi_m^{\{s_n\}}}\,.
\end{equation}

As we discussed in the main text, there are two steps that we also consider separately: (a) to move an excitation from the source to the target using Quantum Zeno Dynamics (which is analogue to the one discussed in Section \ref{sec:heralded0}); (b) herald the transfer by inducing a change in the detector atom state. The second step has to be done fast in order to get the desired scaling of infidelity. We will consider these processes separately. We assume the initial state of the detector is $\ket{s}_d$ such that it only participates in step (b).

\subsubsection{Step (a): From source to target using Zeno dynamics (no-jump evolution).}

\begin{figure}[bt]
	\centering
	\includegraphics[width=0.95\textwidth]{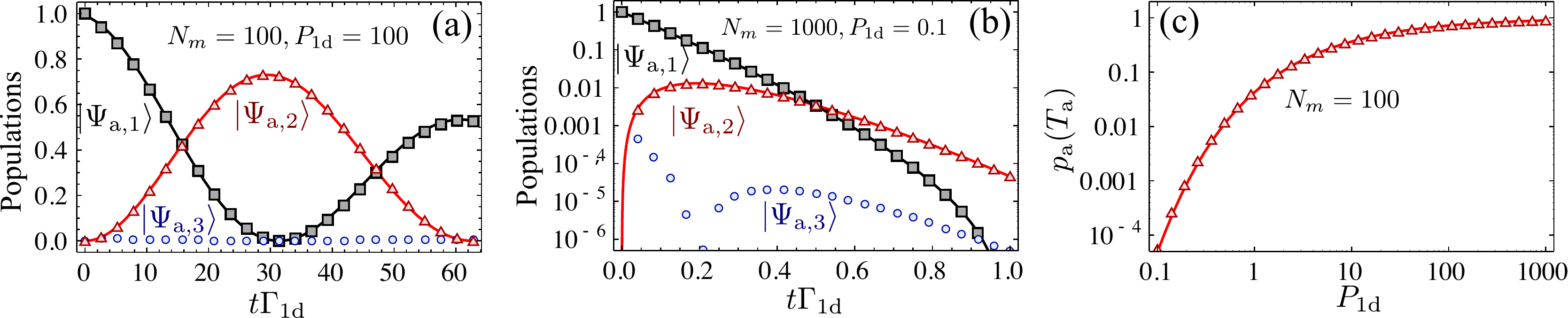}
	\caption{(a-b): Evolution of populations calculated with exact numerical integration (markers), together with the analytical approximations (solid lines, see text) by choosing the optimal $\Omega_\aa=\sqrt{N_m\Gamma_\oned\Gamma^*}$. The two panels correspond to a situation with $N_m=100=P_\oned$ (a) and $N_m=1000$, $P_\oned=0.1$ (b). (c) Exact (solid lines) and analytical optimal probability using $T_\aa\approx \pi/\sqrt{\Gamma^*\Gamma_\oned}$.}\label{figSM2}
\end{figure}

This first step is exactly analogue to the first step of the second protocol discussed in Section \ref{sec:heralded0}. We start in an initial state  $\ket{\Psi_{0,\aa}}=\ket{e}\otimes\ket{\Phi_m^{\{s_n\}}}=\ket{\Psi_{\aa,1}}$, which couples to other states, namely, $\ket{\Psi_{\aa,2}}=\ket{g}\otimes \frac{S_{e_1 g}}{\sqrt{N_m}}\ket{\Phi_m^{\{s_n\}}}$ and $\ket{\Psi_{\aa,3}}=\ket{g}\otimes \frac{S_{a_1 g}}{\sqrt{N_m}}\ket{\Phi_m^{\{s_n\}}}$. For completeness, we compare the analytical expression of the dynamics obtained in Section \ref{sec:heralded0} with the full integration of master equation in Figs.~\ref{figSM2}(a-b), where we observe that they are a good approximation even for systems with $P_\oned<1$, as long as $N_m P_\oned$. The optimal heralding probability
\begin{align}
\label{eqSM:proba}
p_\aa \approx \frac{N_m}{N_m+1} e^{-\pi/\sqrt{P_\oned}}\,, 
\end{align}
that we compare to the exact result in Fig.~\ref{figSM2}(c), being a good approximation even when $P_\oned <1$.

\subsubsection{Step (b): heralding the transfer  using fast $\pi$-pulses (no-jump evolution).}

This second step of the protocol do not rely on quantum Zeno dynamics, and it will take place between the target and detector atoms. As we explained in the main text the idea is to: i) first do a fast $\pi$ pulse with $\Omega_{b}^\rt\gg \Gamma_{\oned}$ such that the possible excitation in $S_{a_1 g}\ket{\Phi_m^{\{s_n\}}}\rightarrow S_{e_2 g}\ket{\Phi_m^{\{s_n\}}}$; ii) let the system evolve only through the couplings induced by the waveguide; iii) do a $\pi$ pulse in both the target and detector ensemble, i.e., $\Omega^\rt_{b},\Omega_{b}^\rd\gg \Gamma_{\oned}$, such that any remaining excitation in the excited states $\ket{e_2}$ go back to $\ket{a_1}$. It is important to notice that if no excitation has been transferred to $\ket{a_1}$ in step (a), nothing will happen in this step. Therefore, we consider our initial state to be: $\ket{\Psi_{0,\bb}}=\ket{\Psi_{b,1}}=\frac{1}{\sqrt{N_m}} S_{e_2 g} \ket{\Phi_m^{\{s_n\}}}\otimes \ket{s}$. The effective Hamiltonian in this case is then given by:
\begin{align}
	H_{\bb,\mathrm{eff}} &=-i\frac{\Gamma^*}{2}\sum_{n}\sigma^{n}_{e_2 e_2}-i\frac{\Gamma_{\oned}}{2}\big(S_{e_2 s}+\sigma_{e_2 s}^\rd\big)\big(S_{s e_2}+\sigma_{s e_2}^\rd\big)\,,
 \label{eq:Hameffb}
\end{align}
which connects $\ket{\Psi_{b,1}}$ only to $\ket{\Psi_{b,2}}=\frac{1}{\sqrt{N_m}} S_{sg} \ket{\Phi_m^{\{s_n\}}}\otimes \ket{e_2}$. In this basis the effective Hamiltonian reads:
\begin{equation}
	H_{\bb,\mathrm{eff}}=
	\frac{1}{2} \left( \begin{array}{ccc} 
	-i (\Gamma^*+\Gamma_{\oned}) & -i\Gamma_{\oned}  \\
	-i \Gamma_{\oned} & -i (\Gamma^*+\Gamma_{\oned})
	\end{array} \right)\,.
\end{equation}

It is again instructive to rewrite it in the basis of sub/superradiant state: $\{\ket{\Psi_{\bb,D}}=\frac{1}{\sqrt{2}} \left( \ket{\Psi_{\bb,1}} - \ket{\Psi_{\bb,2}} \right),\ket{\Psi_{\bb,S}}=\frac{1}{\sqrt{2}} \left( \ket{\Psi_{\bb,1}} + \ket{\Psi_{\bb,2}} \right)\}$, arriving to:
\begin{equation}
	H_{\bb,\mathrm{eff}}=
	\frac{1}{2} \left( \begin{array}{ccc} 
	-i \Gamma^* & 0 \\
	0 & -i (\Gamma^*+2 \Gamma_{\oned})
	\end{array} \right)\,.
\end{equation}
where the Hamiltonian is diagonal. From here, it is simple to arrive to the solution of the dynamics: $\ket{\Psi_\bb (t)}=e^{-i H_{\bb,\mathrm{eff}} t}\ket{\Psi_{0,\bb}}=\sum_{j}\beta_j(t)\ket{\Psi_{\bb,j}} $, which read:
\begin{align}
  \label{eqSM:probabilities2}
 |\beta_1(t)|^2 &=\frac{1}{4}e^{-\Gamma^* t} \Big[ 1+e^{-\Gamma_\oned t}\Big]^2\,\\
  |\beta_2(t)|^2 &=\frac{1}{4}e^{-\Gamma^* t} \Big[ 1-e^{-\Gamma_\oned t}\Big]^2\equiv p_\bb \,,
\end{align}
in the original basis. Notice that $p_\bb$ will be the probability of having transferred the collective excitation to the target while changing the detector atom state to $a_1$, if we assume that the second $\pi$ pulse is perfect. If we choose a time $T_\bb=1/\Gamma_{\oned}$, then
\begin{align}
  \label{eqSM:probabilities2}
   |\beta_1(T_\bb)|^2 & \approx \frac{(e+1)^2}{4e^2}\big(1-\frac{1}{P_\oned}\big)=0.46\big(1-\frac{1}{P_\oned}\big)\,,\\
 p_\bb & \approx \frac{(e-1)^2}{4e^2}\big(1-\frac{1}{P_\oned}\big) \approx 0.1\big(1-\frac{1}{P_\oned}\big)\,,
\end{align}

Interestingly, there is a sizeable probability of remaining in the initial state of this step. So instead of start from the beginning every time that we fail, it is possible to repeat this step several times, which increases probability to $p_b\approx 1/3$.

\subsubsection{Step a: Using Zeno dynamics (quantum jump analysis).}

As we already calculated in Section \ref{sec:heralded0}, the probability of having an individual quantum jump in the target ensemble is given by:
 \begin{align}
  \label{eqSM:summ}
  &p_{\aa,*}\lesssim \frac{\pi}{(N_m+1)\sqrt{P_\oned}}\,,\,\mathrm{if }\,\,\,\,P_\oned\gg 1\\
  & p_{\aa,*}\lesssim \frac{3}{(N_m+1)}\,,\,\mathrm{if }\,\,\,\, P_\oned\ll 1\ \,.
 \end{align}

\begin{figure}[bt]
	\centering
	\includegraphics[width=0.75\textwidth]{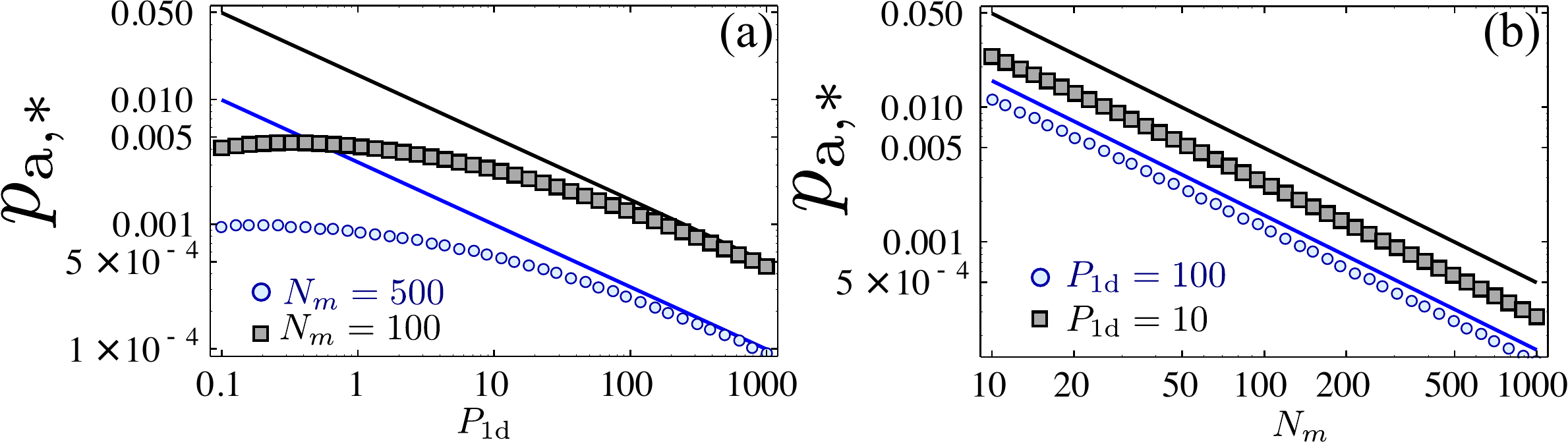}
	\caption{(a-b): Exact (markers) vs estimated upper bound (solid lines) $p_{\aa,*}$ as a function of $P_\oned$ (a) and $N_m$ (b). The exact calculation is obtained from integrating populations in Eq.~\ref{eqSM:probabilities2}, whereas for the analytical upper bound we use the leading order contribution: $\frac{\pi}{2(N_m+1)\sqrt{P_\oned}}$. The $T_\aa$ is chosen to be the optimal one as in Fig.~\ref{figSM2}.}
	\label{figSM3}
\end{figure}

  In Fig.~\ref{figSM3} we compare the exact results (markers) with the upper bound (solid lines) for $\varepsilon_a$, showing that actually we get the right scalings both with $P_\oned$ and $N_m$.
  
\subsubsection{Step b: heralding the transfer using fast $\pi$-pulses (quantum jump analysis).}

The relevant quantum jumps in this step are $J_{\bb}\rho=J_{\bb,*}\rho+J_{\bb,\mathrm{coll}}\rho$, with
 \begin{align}
  \label{eq:qjumpstepb}
  J_{\bb,*}\rho&=\frac{\Gamma^*}{2} \sum_{n}\sigma^{n}_{s e_2}\rho\sigma^{n}_{e_2 s}+\frac{\Gamma^*}{2}\sum_{n}\sigma^{n}_{a_1 e_2}\rho\sigma^{n}_{e_2 a_1}\,,\nonumber\\
  J_{\bb,\mathrm{coll}}\rho&= \Gamma_{\oned}\big(S_{s e_2}+\sigma_{s e_2}^\rd\big)\rho\big(S_{e_2 s}+\sigma_{e_2 s}^\rd\big)\,.
 \end{align}

In this second step, we make a $\pi$-pulse with $\Omega_{b}^\rt$, such that the initial state $\ket{\Psi_{\bb,0}}=\ket{\Psi_{\bb,1}}=\frac{1}{\sqrt{N_m}}S_{e_2 g}\ket{\Phi_m^{\{s_n\}_n}}$. Then the system is left free to evolve for a time $T_\bb=\frac{1}{\Gamma_\oned}$, and switching the $\Omega_{b}^\rt,\Omega_{b}^\rd$ such that no population remains in the excited state $e_2$. Therefore, the quantum jumps could only occur within a small time interval giving rise to
 \begin{align}
  \label{eqSM:pbad2}
 p_{\bb,*}=\Gamma^*\int_0^{T_\bb} dt_1 |\beta_1|^2\approx \frac{0.67}{P_\oned}\,.
 \end{align}

\begin{figure}[bt]
	\centering
	\includegraphics[width=0.95\textwidth]{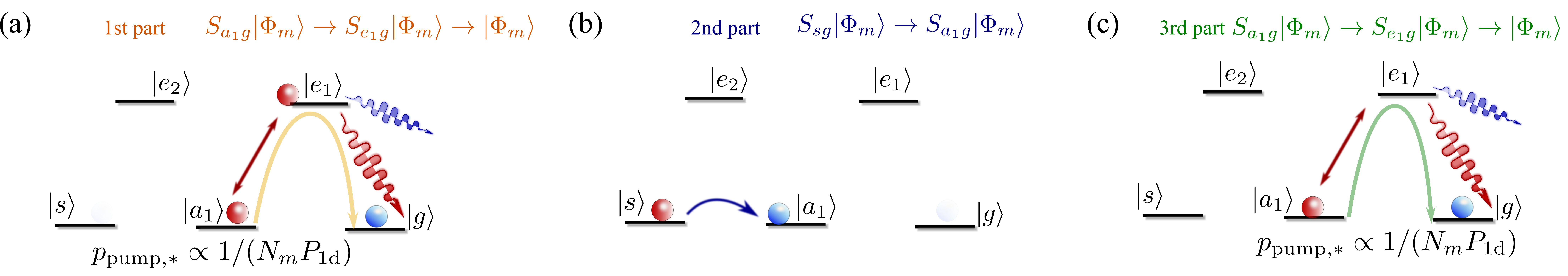}
	\caption{General scheme of repumping process to correct collective quantum jump errors (see details in the text).}
	\label{figSM4}
\end{figure}

 In this regime, with $P_\oned\gg 1$ and $T_\bb=\frac{1}{\Gamma_\oned}$, it is also instructive to consider the probability of other quantum jumps. For example, to calculate the probability that the excitation is transferred to $S_{sg}\ket{\Phi_m^{\{s_n\}}}$ but the state of the detector is unchanged, i.e., remains in $s$, one needs to sum up the probability of making a collective quantum jump in the target together with the probability of making any jump in the $f\rightarrow s$ transition:
 \begin{align}
  \label{eqSM:pbad2col}
p_{\bb,\mathrm{coll}}=\Gamma_\oned \int_0^{T_\bb} dt_1 |\beta_1|^2+(\Gamma_\oned+\frac{\Gamma^*}{2}) \int_0^{T_\bb} dt_1 |\beta_2|^2\approx 0.71\,, 
 \end{align}
which biggest contribution comes from the emission in the target ensemble. These states have not destroyed the coherence of $\ket{\Phi_m^{\{s_n\}}}$ but they need to be properly taken care of with the repumping process. In order to avoid that these states also lead to error, we have to make a repumping process back to $\ket{g}$ in the way depicted in Fig.~\ref{figSM4}:
\begin{enumerate}
\item First we pump incoherently back any possible excitation in $\ket{c}$ through the waveguide, namely, $S_{a_1 g}\ket{\Phi_m^{\{s_n\}}}\rightarrow S_{e_1 g}\ket{\Phi_m^{\{s_n\}}}\rightarrow\ket{\Phi_m^{\{s_n\}}}$. Because this process is done through a collective photon the error introduced in this step $p_{\mathrm{pump},*} \propto \frac{1}{NP_\oned}$.
\item Once, we have make sure that there are no excitation in $\ket{a_1}$ we can move the excitations from $S_{sg}\ket{\Phi_m^{\{s_n\}}}\rightarrow S_{a_1 g}\ket{\Phi_m^{\{s_n\}}}$ with $\pi$ pulse using a microwave or two-photon Raman transition. 
\item Now, after having transferred the excitation to $\ket{a_1}$, we repeat the incoherent transfer through the waveguide that will only introduce errors of the order $p_{\mathrm{pump},*}$.
\end{enumerate}

\subsection{Fidelities of the protocol.}

From the calculation of probabilities of the previous Section, we are prepared to estimate the average fidelity when detecting an atomic excitation in the detector atom in $a_1$. The errors appear:
\begin{itemize}
\item \emph{Successful heralding:} As shown in the main manuscript, in the case where the transition $e_1\rightarrow\ket{a_1}$ is closed the only way of measuring detector atom in $\ket{a_1}$ is by successfully generating $\ket{\Phi_{\mathrm{goal}}}$. If we have a spurious spontaneous emission $\sim \alpha \Gamma^*$, then, it is possible that we herald the transfer of an incoherent excitation of $s$, i.e., $\sigma_{sg}^n$, which have an overlap $\propto 1/N$ with the state we want to create. Thus, the possible error after heralding is given by:
\begin{equation}
\label{eqerroclosed}
\varepsilon_{\mathrm{closed}}=\frac{\alpha}{N\sqrt{P_\oned}}\,.
\end{equation}

\item \emph{Failed heralding:} As it occurred in the first protocol, in each failed attempt we introduce an error in our initial state $\ket{\Phi_m^{\{s_n\}}}$ due to the probability of emitting free space, which in this protocol may occur in step (a), (b) and the repumping. The error per attempt can be estimated to be:
\begin{equation}
\label{eqerrojump}
\varepsilon_{*}\approx (p_{\aa,*}+p_{\bb,*}+p_{\mathrm{pump},*})\frac{m}{N}\approx \frac{0.67 m}{N P_\oned} \,.
\end{equation}
where in the last approximation we assumed to be in a regime where $p_{\bb,*}\gg p_{\aa,*},p_{\mathrm{pump},*}$.

\item \emph{Complete process:} As in the first protocol, the average complete process consist of a successful heralding and $1/p$ failed attempts, such that the average fidelities of the process:
\begin{align}
\label{eqSM:fid11}
I_{m\rightarrow m+1}=\varepsilon_{\mathrm{closed}}+\frac{\varepsilon_{*}}{p}=\frac{\alpha}{N\sqrt{P_\oned}}+\frac{ m}{p N P_{\oned}}\,,
\end{align}
 where one sees that to reach the $I_{m->m+1}\propto 1/(N P_\oned)$ scaling, we do not need a perfectly closed transition, but it is enough to demand a cancellation of spontaneous emission ($\alpha$) of at least $\alpha=1/\sqrt{P_\oned}$ which can be achieved as well, e.g., by using a quadrupole transition.

\end{itemize}

\begin{figure}[bt]
	\centering
	\includegraphics[width=0.75\textwidth]{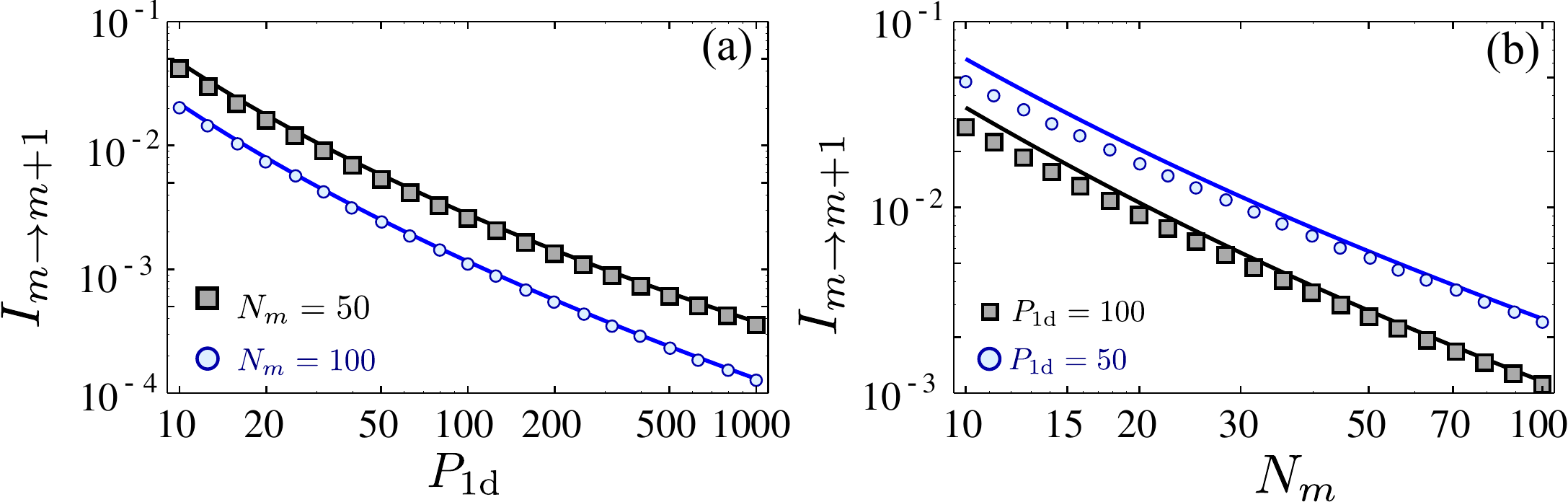}
	\caption{Scaling of the exact (solid lines) and approximated (markers) infidelities with $P_\oned$ (Panel a) and $N_m$ (Panel (b)) for different $N_m$ and $P_\oned$ respectively as depicted in the legends. }
	\label{figSM5}
\end{figure}

 In Fig.~\ref{figSM5}, we compare the analytical approximation of infidelities of Eq.~\ref{eqSM:fid11} (markers) together with exact results (solid lines) as a function of $P_\oned$ and $N_m$.

 \section{Heralding single photon excitations using atom-waveguide QED: single-step protocol for polynomial scaling.\label{sec:heralded2}}

In this Section, we study a modification of the previous protocol such that we can do the heralded transfer in a single step and without the need of the closed transition at expense of showing a worse scaling of the success probabilities and infidelities and with the requirement of two different guided modes. These protocol is not discussed in the main manuscript, though we include here for completeness.

 \subsection{Atom waveguide resources.}
 
\begin{figure}[tb]
	\centering
	\includegraphics[width=0.35\textwidth]{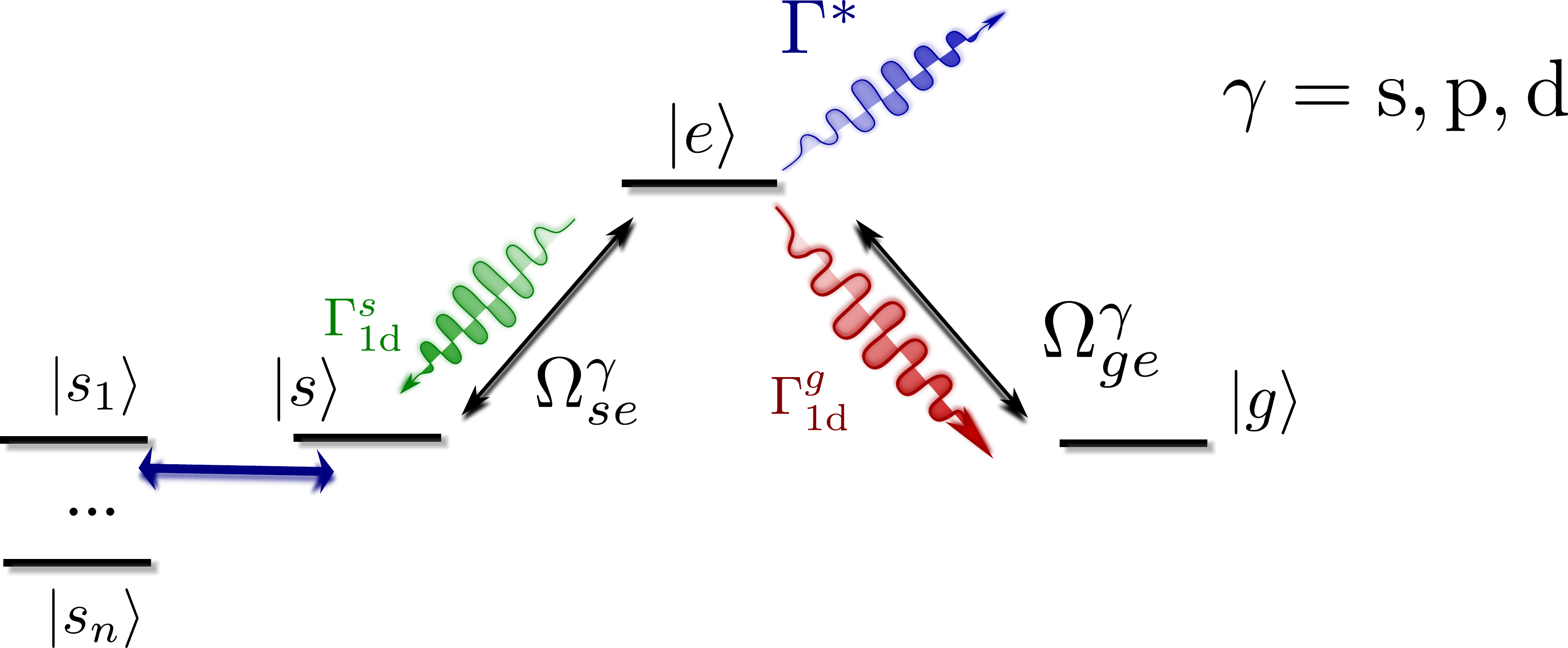}
	\caption{Variation of the internal level structure of emitters of the different ensembles in which the transition $\ket{e}\leftrightarrow\ket{g}$ [$\ket{e}\leftrightarrow \ket{g}$ is coupled to a single waveguide mode with $\Gamma_{\oned}$. We also require an extra intermediate level $\ket{c}$, which forms a closed transition with $\ket{e}$. We also assume that we can control the $\ket{e}\leftrightarrow \ket{g}[\ket{s}] $ transition with a classical laser $\Omega_{ge}^{\alpha}$ [$\Omega_{se}^{\alpha}$].  }\label{figSM6}
\end{figure}

 The requirements for this protocol are:
 \begin{itemize}
 \item We need to have an atomic configuration of source/target/detector atom as in the previous protocol. However, the internal level structure must be modified, as shown in Fig.~\ref{figSM6} to have an excited state $\ket{e}$ coupled directly to two waveguide modes through the transitions $\ket{e}\leftrightarrow\ket{g[,s]}$ with rates $\Gamma_{\oned}^{g [,s]}$.  As we will see afterwards in this case it is important that $\Gamma_{\oned}^g\neq \Gamma_{\oned}^s$. The modes can differ either in polarization or frequency, but they must satisfy that their characteristic wavelength $\lambda_\aa$ is approximately the same.
 
 \item To guarantee the heralding, it will be important that the transition $\ket{e}\rightarrow \ket{s}$ [$\ket{e}\rightarrow \ket{g}$] of the source [detector] atom to be largely off-resonant. This must be satisfied to avoid a direct transfer of the excitation from the source to the detector atom that would spoil the heralding.
 \end{itemize}

\subsection{Protocol and calculation of probabilities.}

The main difference with the previous protocol is that is done in a single step. We start with  $\ket{\Psi_{0,\aa}}=\ket{\Psi_{\aa,1}}=\ket{e}\otimes\ket{\Phi_m^{\{s_n\}}}\otimes\ket{s}$ and consider that only the classical field $\Omega_{ge}^\rd\equiv \Omega^\rd\neq 0$, such that the dynamics governed by the effective Hamiltonian:
\begin{align}
	H_{\aa,\mathrm{eff}} &= \frac{\Omega^\rd}{2}\big(\sigma_{ge}^{\rd}+\sigma_{eg}^{\rd}\big)-i\frac{\Gamma^*}{2}\sum_{n}\sigma^{n}_{ee}-i\frac{\Gamma_{\oned}^\rg}{2}\big(\sigma_{eg}^\rs+ S_{eg}\big)\big(\sigma_{ge}^\rs+S_{ge}\big)-i\frac{\Gamma_{\oned}^\rs}{2}\big(\sigma_{es}^\rd+S_{es}\big)\big(\sigma_{se}^\rd+ S_{se}\big)\,,
 \label{eqSM:Hameff}
\end{align}
that couples our initial state to $\ket{\Psi_{\aa,2}}=\ket{g}\otimes\frac{1}{\sqrt{N_m}}S_{eg}\ket{\Phi_m^{\{s_n\}}}\otimes\ket{s}$, $\ket{\Psi_{\aa,3}}=\ket{g}\otimes\frac{1}{\sqrt{N_m}}S_{sg}\ket{\Phi_m^{\{s_n\}}}\otimes\ket{e}$ and $\ket{\Psi_{\aa,4}}=\ket{g}\otimes\frac{1}{\sqrt{N_m}}S_{sg}\ket{\Phi_m^{\{s_n\}}}\otimes\ket{g}$. In this basis the Hamiltonian can be written as:
\begin{equation}
	H_{\mathrm{eff}} = 
	\frac{1}{2} \left( \begin{array}{cccc} 
	-i (\Gamma_\oned^\rg+\Gamma^*) & -i \sqrt{N_m} \Gamma_\oned^\rg & 0  & 0 \\
	-i \sqrt{N_m}\Gamma_\oned^\rg  & - i (N_m \Gamma_\oned^\rg+\Gamma_\oned^\rs+\Gamma^*) & -i \Gamma_\oned^\rs  & 0 \\
	0 & -i \Gamma_\oned^\rs & -i (\Gamma_\oned^\rs+\Gamma^*) & \Omega^\rd \\
	0 & 0 & \Omega^\rd & 0
	\end{array} \right)\,,
	\label{eq:Ham1step}
\end{equation}

 The goal is to find $\ket{\Psi(t)}=e^{-i H_{\aa,\mathrm{eff}} t}\ket{\Psi_0}=\sum_{j}\alpha_j(t)\ket{\Psi_j}$, which can always be done numerically as it is only a 4x4 matrix. However, in order to gain more insight it is useful to write it in a basis with sub and superradiant states. Interestingly, the system always contains a dark state $\ket{\Psi_{\aa,D}}=\frac{1}{\sqrt{N_m+2}} \left( \sqrt{N_m} \ket{\Psi_{\aa,1}}-\ket{\Psi_{\aa,2}}+\ket{\Psi_{\aa,3}}\right)$, which is dark for all values of $\Gamma_\oned^\rg,\Gamma_\oned^\rs \neq 0 $. However, the two orthogonal excited states depend on the ratio $\Gamma_\oned^\rs / \Gamma_\oned^\rg$. After a careful study, we found the optimal choice for the maximum transfer to $\ket{\Psi_{\aa,4}}$ was to choose  $\Gamma_\oned^\rs / \Gamma_\oned^\rg = \frac{N_m+1}{2}$, and the reason is that in that case the two decay channels to $\ket{g}$ and $\ket{s}$ show the same "superradiant" decay $(N_m+1)\Gamma_\oned^\rg$.  From now one, we use that ratio and $\Gamma_{\oned}
^g=\Gamma_{\oned}$. For this ratio, the adiabatic elimination of the superradiant states give rise to an effective $2\times 2$ between the dark state and the goal state
\begin{equation}\label{eqSM:eff1step}
H_{\mathrm{eff}} \approx \left(
\begin{array}{cc}
 -\frac{i \Gamma^*}{2} &  \frac{\Omega^\rd}{2\sqrt{N_m}} \\
   \frac{\Omega^\rd}{2\sqrt{N_m}} & -i\frac{3  (\Omega^\rd)^2 }{2N_m\Gamma_\oned}\\
\end{array}
\right)\,.
\end{equation}
where we have used $N_m\gg 1$ to simplify the expressions. As we know from the previous Sections, the more convenient is to choose an $\Omega^\rd$ such that the contribution of $\Gamma^*$ and the effective losses induced by superradiant states, i.e., $\frac{3(\Omega^\rd)^2 }{2N_m\Gamma_\oned}$, is the same. Thus, by choosing $\Omega^\rd = \sqrt{N_m \Gamma_\oned \Gamma^* / 3}$, it is easy to find that:
\begin{align}
\label{eqSM:a}
|\alpha_D(t)|^2 &\approx \frac{N_m}{N_m+2} e^{-\Gamma^*t}\cos^2(\frac{\sqrt{\Gamma_\oned\Gamma^*}t}{2\sqrt{3}})\,,\nonumber \\
|\alpha_4(t)|^2&\approx \frac{N_m}{N_m+2} e^{-\Gamma^*t}\sin^2(\frac{\sqrt{\Gamma_\oned\Gamma^*}t}{2\sqrt{3}})=p(t)\,.
\end{align}

Therefore, if $P_\oned\gg1$ and $T_\aa=\frac{\pi\sqrt{3}}{\sqrt{\Gamma_{\oned}\Gamma^*}}$ then the probability of heralding reads:
\begin{align}
\label{eqSM:pbig}
p\approx \frac{N_m}{N_m+2} e^{-\sqrt{3}\pi/\sqrt{P_\oned}}\,,
\end{align}

\begin{figure}[tb]
	\centering
	\includegraphics[width=0.89\textwidth]{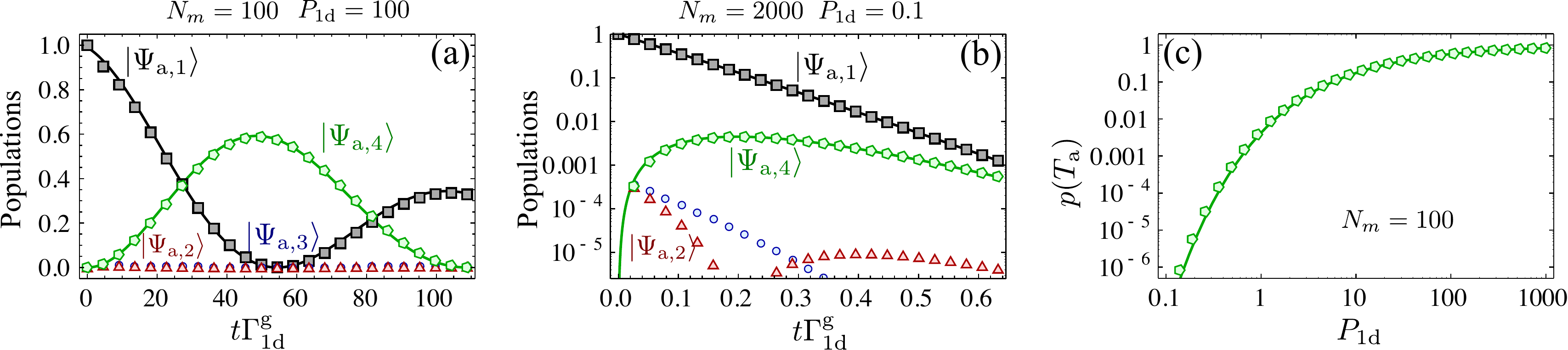}
	\caption{(a-b): Evolution of populations calculated with exact non-hermitian hamiltonian (markers) of Eq.~\ref{eq:Ham1step}, together with the analytical approximations (solid lines) by using the superradiant/subradiant decoupling of Eq.~\ref{eqSM:eff1step} by choosing the optimal $\Omega_\aa=\sqrt{N_m\Gamma_\oned\Gamma^*/3}$. The two panels correspond to a situation with $N_m=100=P_\oned$ (a) and $N_m=2000$, $P_\oned=0.1$ (b). (c) Exact (solid lines) and analytical optimal probability using $T_\aa\approx \pi\sqrt{3}/\sqrt{\Gamma^*\Gamma_\oned}$.}\label{figSM7}
\end{figure}

In Fig.~\ref{figSM7} we compare the results from the exact integration of the complete non-hermitian hamiltonian together with the analytical approximation (markers) obtained from the Quantum Zeno dynamics (solid lines).

The problematic quantum jump in this case is related to the probability of having an excitation $\ket{e}$ in the \emph{target} ensemble, i.e., coming from the population of state $\ket{\Psi_{\aa,2}}$ that reads
\begin{align}
  \label{eqSM:pbad11}
  p_{*}=\Gamma^*\int_0^{T_\aa} dt_1 |\alpha_2(t_1)|^2+ \Gamma^*\int_{0}^\infty d t_1|\alpha_2(T_\aa)|^2 e^{-\Gamma^* t_1}\,,
 \end{align}
where the first part corresponds to the interval of time $(0,T_\aa)$ where $\Omega^\rd$ is switched on, and the second part, the time $t\gg 1/\Gamma^*$ that we wait, with  $\Omega^\rd=0$, such that all the population in the dark state vanishes. The contributions of the superradiant/subradiant states to $\ket{\Psi_{\aa,2}}$ can be obtained in the asymptotic limit $N_m\gg 1$, where:
\begin{align}
\label{eqSM:overlap}
|\braket{\Psi_{\aa,D}}{\Psi_{\aa,2}}|&\sim \frac{1}{\sqrt{N_m}}\,\nonumber\\
|\braket{\Psi_{\aa,S_{+}}}{\Psi_{\aa,2}}|&\sim \frac{\sqrt{2+\sqrt{2}}}{2}\sim O(1)\,\nonumber\\
|\braket{\Psi_{\aa,S_{-}}}{\Psi_{\aa,2}}|&\sim \frac{\sqrt{2-\sqrt{2}}}{2}\sim O(1)\,\nonumber\\
\end{align}

\begin{figure}[bt]
	\centering
	\includegraphics[width=0.75\textwidth]{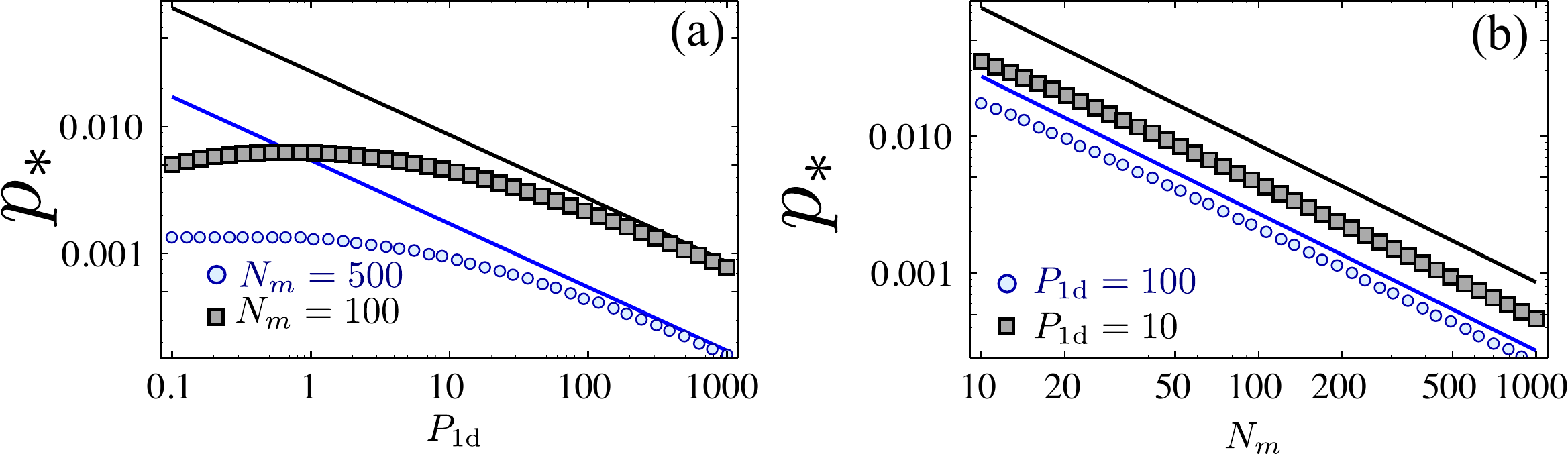}
	\caption{(a-b): Exact (markers) vs estimated upper bound (solid lines) $p_{*}$ as a function of $P_\oned$ (a) and $N_m$ (b). The exact calculation is obtained from integrating populations in Eq.~\ref{eqSM:pbad11}, whereas for the analytical upper bound we use the leading order contribution of Eq.~\ref{eq:pbad11sec}. The $T_\aa$ is chosen to be the optimal one as in Fig.~\ref{figSM7}.}
	\label{figSM8}
\end{figure}

Using that information we can estimate the contribution of the dark state within the time interval: $(0,T_\aa)$.
\begin{align}
  \label{eq:pbad11}
 p_{*,D}\lesssim\frac{\Gamma^*}{N_m}\int_0^{T_\aa} dt_1 e^{-\Gamma^* t_1}\lesssim  \frac{1-e^{-\Gamma^* T_\aa}}{N_m}\,,
 \end{align}
that is:
\begin{align}
  \label{eq:pbad11sec}
  p_{*,D}&\approx \frac{\pi \sqrt{3}}{2 N_m\sqrt{P_\oned}}\,,\,\mathrm{if}\,\,\,, P_\oned\gg 1\,,
    \end{align}
 whereas the contribution of $(T_\aa,\infty)$ can always be shown to be smaller order: when $P_\oned\gg 1$. Therefore, we the overall probability of emitting a quantum jump in the target ensemble for this protocol 
\begin{align}
  \label{eq:pbad11sec}
  p_{*}&\propto \frac{\pi\sqrt{3}}{2N_m\sqrt{P_\oned}}\,,\,\mathrm{if}\,\,\,, P_\oned\gg 1\,,
  \end{align}
 which are confirmed numerically in Fig.~\ref{figSM8}. The repumping process to correct the collective quantum jump errors can be done as well in a single step by pumping collectively the atoms from $\ket{s}\rightarrow \ket{g}$, which moves $S_{sg}\ket{\Phi_m}\rightarrow S_{eg}\ket{\Phi_m}\rightarrow\ket{\Phi_m}$.  Because this process is done through a collective photon the probability of emitting a free space photon is given by $p_{\mathrm{pump},*} \propto \frac{1}{NP_\oned}$.

\subsection{Fidelity of the protocol.}

\begin{figure}[bt]
	\centering
	\includegraphics[width=0.75\textwidth]{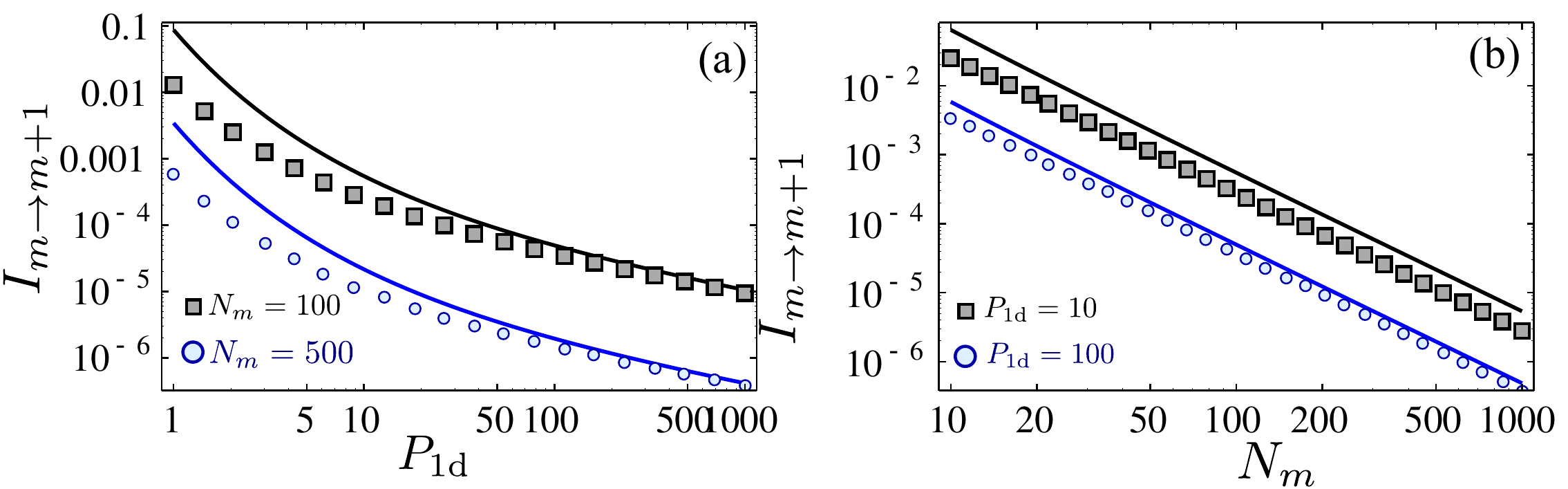}
	\caption{Scaling of the exact (solid lines) and approximated (markers) infidelities with $P_\oned$ (Panel a) and $N_m$ (Panel (b)) for different $N_m$ and $P_\oned$ respectively as depicted in the legends. }
	\label{figSM9}
\end{figure}

The analysis of the fidelity will be very similar to the one performed for the previous protocol. In this case, if we detect an excitation in $g$ in the detector atom, no error may appear. Therefore, the only contribution to the infidelity is the error coming from free space photons in each attempt after $1/p$ repetitions, which finally scale:
\begin{equation}
\label{eq:p}
I_{m\rightarrow m+1}\approx \frac{1}{p}\times (p_{*}+p_{\mathrm{pump},*})\times\frac{m}{N}\propto \frac{m}{N^2 \sqrt{P_\oned}}\,,
\end{equation}
for systems with $P_\oned,N_m\gg 1$. In Fig.~\ref{figSM9} we confirm numerically that the analytical approximations capture the right scaling. If we consider as a reference the largest Purcell factor (the one associated to $\Gamma_{\oned}^\rs$), then, the scaling of probabilities and fidelities should read equivalently:
\begin{align}
\label{eq:fidsinglestep}
p&\approx \frac{N_m}{N_m+2} e^{-\sqrt{3}\pi \sqrt{N_m}/\sqrt{P_\oned}}\,\nonumber \\
I_{m\rightarrow m+1}&\propto \frac{m}{N^{3/2} \sqrt{P_\oned}}\,,
\end{align}

\section{Merging protocols for combining collective atomic excitations.\label{sec:doubling}}

In this Section we show how to combine the excitations stored in the target atoms after the first part of the process, that is, when our atomic state has the form $\ket{\Phi_m^{\{s_n\}}}\propto \mathrm{sym}\{\ket{s_1}^{m_1}\otimes \dots \otimes\ket{s_n}^{m_n}\otimes \ket{g}^{N_m}\}$, with $N_m=N-m=N-\sum_{i} m_i$ such that we can build up higher excitation numbers into a single level. For that purposes we use several tools and assumptions:
\begin{itemize}
\item We are interested in the so called Holstein-Primakoff limit where $m\ll N$, such that the collective atomic dipoles $S_{\alpha g}\sqrt{N_m}\approx a^\dagger_\alpha$  can be approximated by bosonic operators. 
\item Using the microwave/Raman lasers connecting an atomic level $\alpha$ with $g$, we can do displacement transformations $D(\alpha)$ on the $\alpha$ level. Moreover, using microwave/Raman transition between $\alpha$ and $\beta$ levels, we can also engineer beam splitter transformations between these modes.
\item Moreover, we can read the atomic state very efficiently by pumping to an excited state that emits a collective photon through the waveguide in a cyclic transition. This will allow do to very efficient detection and potentially number resolved as we explain in the last Section. Together with the beam splitter and displacement transformations, they provide a similar set of tools as the ones used in linear optics protocol with the advantage of having the excitations stored. 
\end{itemize}

Using all those assumptions, our goal is to build a given arbitrary atomic state of $m$ excitations, that we know it can be afterwards be mapped to a photonic state with very high efficiency \cite{gonzaleztudela15a}. The general form of this state typically reads:
\begin{align}
\label{eqSM:goal}
\ket{\Psi_{m}}=\sum_n \frac{f_n}{\sqrt{n!}}(a^\dagger)^n\ket{\mathrm{vac}}\propto \prod_{i=1}^n(a^\dagger-\alpha_i^*)\ket{\mathrm{vac}}\,,
\end{align}
and we will put particular emphasis in the simple Dicke states with $m$ excitations, as this will be mapped to photonic Fock states. We discuss two different types of merging protocols separately.
\begin{enumerate}
\item If the number of metastable states $\ket{s_i}$ is limited, we only use two levels and add excitations one by one. This is the kind of protocols typically used in linear optics protocols where it is not possible to store several atomic excitations \cite{dakna99a}.
\item Using several hyperfine levels and adopting a tree-like structure that will allow us to avoid the exponential scaling of the number of operations.
\end{enumerate}

\subsection{Adding excitations one by one.}

\begin{figure}[tb]
	\centering
	\includegraphics[width=0.97\textwidth]{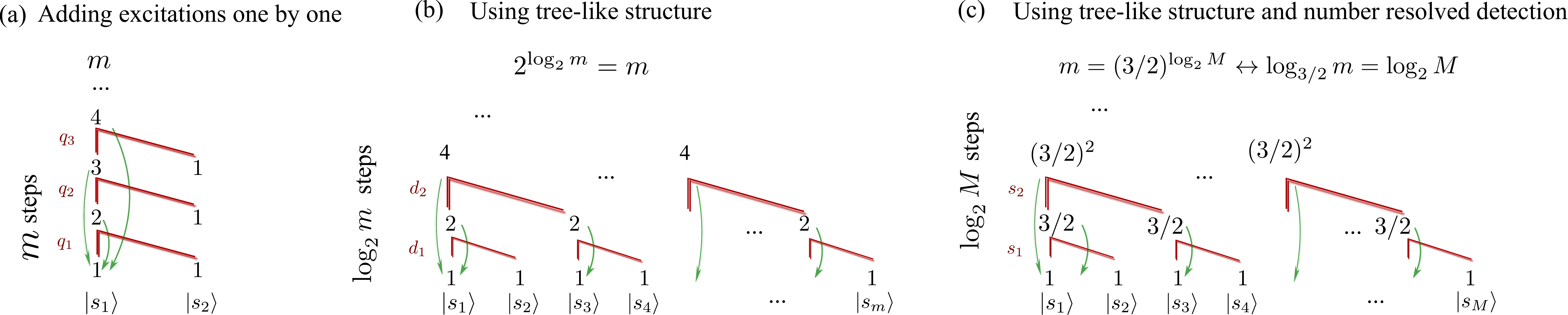}
	\caption{Scheme to build up $m$ collective atomic states combining heralded single collective excitations plus beam splitters and postselection. The red arrows indicates the path adding excitations, whereas the green ones how one needs to reinitialize the process if we fail in given step. (a) Scheme using two atomic levels to add excitations one by one with probability $q_n$ in each step.  (b) [(c)] Scheme using $\log_2 m$ [$\log_{3/2} m$] with probability $q_n$ [$s_n$] in each step using a tree-like structure that avoids starting the whole process from the beginning in case of failure. }\label{figSM10}
\end{figure}

Here we show how to create a photonic state of $m$ excitations by combining beam splitters ($B$), displacements transformations ($D(\alpha)$) and post-selection condition on no detection in a particular level $\mathbb{P}_{0_i}$ by using only two atomic levels. For the sake of discussion, we analyze first in detail the case of single Dicke states and leave the analysis of the superpositions afterwards.

The general scheme to build a given atomic state with $m$ collective excitations is depicted in Fig.~\ref{figSM10}(a). The main step consists of going from an state with $m$ excitations in one level and a single one in the other, $\ket{\phi_n}=\ket{n}\otimes\ket{1}$, to an state $n+1$ excitation in one of them conditioned on detecting no excitation in the other one, i.e.,  $\ket{n+1}\otimes\ket{0}_1$. The idea is to apply a beam splitter transformation of $\ket{\phi_n}$, followed by a projection of the state conditioned on detecting no excitations in the $a_2$ level:
\begin{align}
\ket{\tilde{\phi}_n}=\mathbb{P}_{0_2}B\ket{\phi_n}= \frac{1}{\sqrt{n!}}T^{n} R (a_1^\dagger)^{n+1} \ket{\mathrm{vac}}
\end{align}
(not normalized) and the probability of finding it is given by
\begin{align}
\bra{\tilde{\phi}_n}\tilde{\phi}_n\rangle=(n+1)|T|^{2n}|R|^{2} \leq \big(\frac{n}{n+1}\big)^m =q_{n},
\end{align}
which is maximized for a transmitivity $|T|^{2}=n/(n+1)$. Interestingly, $\lim_{n\rightarrow \infty} q_{n}=e^{-1}$ such that it does not decay with $n$. However, if we fail, we have to repeat the process from the beginning, such that the mean number of operations to create a state with $m$ excitations can be obtained:
\begin{align}
\label{eq:outm3}
R_{m}=\frac{1+R_{m-1}}{q_{m-1}}=\frac{1}{q_{m-1}}+\frac{1}{q_{m-1}q_{m-2}}+\dots +\frac{1}{\prod_{n}^{m-1} q_n}\,.
\end{align}

Taking into account that $2=q_{1}>q_{2}>\dots>q_{n}>e^{-1}$ and $q_0=p$ the probability to generate a single heralded excitation, the mean number of states can be lower and upper bounded by:
\begin{align}
R_m<\frac{m}{\prod_{n}^{m-1} q_n}< \frac{m}{\prod_{n}^{m-1} q_n}< \frac{m e^{m}}{p}\,,\\ \nonumber 
R_m>\frac{1}{\prod_{n}^{m-1} q_n}>\frac{1}{p \prod_{n=1}^{m-1} q_1}>\frac{2^{m-1}}{p}\,.
\end{align}
which increase exponentially to increase the number of excitations $m$. Moreover, it was shown in Ref.~\cite{dakna99a} that by combining single photon addition with displacement transformation before and after the addition, lead to a conditional outcome after starting with a state $\ket{\phi}$, that is, $D(\alpha)a^\dagger D(\alpha)^\dagger\ket{\phi}$. Combining $m$ of these operations and using the fact that  $D(\alpha)a^\dagger D(\alpha)^\dagger=a^\dagger-\alpha$ it can be shown that one arrives to $\ket{\Psi_{\mathrm{ph},m}}$ as in Eq.~\ref{eqSM:goal}, but also with an exponential number of operations.

\subsection{Doubling the number of excitations post-selecting on no detection.}

The key difference with respect to the previous protocol is that we are going to adopt a tree-like structures as depicted in Fig.~\ref{figSM10}(b), where we can reach the $m$ excitations in $\log_2 m$ steps and we do not have to repeat all of them if we fail. The first building block is to study the process that add up two states with $\ket{n}$ excitations to go to $\ket{n}\otimes\ket{n} \rightarrow \ket{2n,0}$. It can be shown that the conditional outcome after a beam splitter and detecting no excitations in the second atomic level is given by:
\begin{align}
\label{eqSN:outmD}
\ket{\tilde{\phi}_{2n}}=\frac{1}{n!}T^{n} R^n (a_1^\dagger)^{2n} \ket{\mathrm{vac}},
\end{align}
which yields an (approximated) optimal probability for a $50-50$ beam splitter transformation:
\begin{align}
\label{eq:P2m}
d_{n}=\braket{\tilde{\phi}_{2n}}{\tilde{\phi}_{2n}}=\frac{(2n)!}{2^{2n}(n!)^2}\approx 1/\sqrt{\pi n}\,,
\end{align}
where we used Stirling's approximation $n! \approx \sqrt{2 \pi n}\ n^n \ee^{-n}$ in the last step. Therefore, in this case the optimal $d_{n}$ does decay with $1/\sqrt{n}$ but only polynomially. However, in spite of this decay, the use of of a tree-like structure is enough to circumvent the exponential scaling of adding excitations one by one. The average number of operations to arrive to an state of $m$ excitations can be calculated iteratively, i.e., 
\begin{equation}
R_m = d^{-1}_{m/2} \left( 1 + 2 R_{m/2} \right)=d^{-1}_{m/2}+2 d^{-1}_{m/2} d^{-1}_{m/4}+2^2 d^{-1}_{m/2} d^{-1}_{m/4}d^{-1}_{m/8}+\dots\,,
\end{equation}
where the $d^{-1}_{m/2}$ terms represents the number of times we need to try in the step $m/2$ to succeed and the one with $2 R_{m/2}$ is the number of steps we need to repeat at the $m$-th step to get the the two branches of the tree. We lower and upper bound this number of operations to get $m$ excitations by:
\begin{align}
R_m &<\frac{\log_2 m}{p} \frac{2^{\log_2 m-1}}{\prod_{n=1}^{\log_2 m-1} d_{m/2^n}}<\frac{\log_2 m}{p}\frac{ 2^{\log_2 m-1}}{d_{m/2}^{\log_2 m}}\approx \frac{ m^{(\log_2 m)/2+1 }\log_2 m}{2 p}\,,\nonumber \\
R_{m} &> \frac{2^{\log_2 m-1}}{p \prod_{n=1}^{\log_2 m-1} d_{m/2^n}}>\frac{m^2}{4 p}\,,
\end{align}
where we use that $d_{m}<d_{m/2}$ and $d_1=1/2$ and that we require $\log_2 m$ steps to arrive to $m$ excitations. Thus, $R_m$ is superpolynomial in $m$ because the probability in each step, $d_m$, decays with $m$. If we had a way of making $d_m$ independent of $m$ as it occurs for the single excitation case $q_m$, then we would have a polynomial scaling. 

For completeness it will be interesting to prove that one can also get arbitrary superpositions \cite{fiurasek05a} using the tree-like structure of Fig.~\ref{figSM10}(b). As a resource we assume we can build any superposition of $0$ and $1$ photon, i.e., which we know is possible from the previous section. Then, we prove first that using $50-50$ beam splitter and detecting no excitations in the $a_2$-mode we can arrive to any arbitrary state as follows:
\begin{align}
\label{eq:gene}
  & P_{0_{2}}B \ket{\Psi_{a_1,m}}\otimes \ket{\Psi_{a_2,m}}
  =P_{0_{2}}B \prod_{i=1}^m\Big(a_1^\dagger-\frac{\alpha_i^*}{\sqrt{2}}\Big)\prod_{i=1}^m\Big(a_2^\dagger-\frac{\alpha_{i+m}^*}{\sqrt{2}}\Big)\ket{\mathrm{vac}}= \nonumber\\
&=\frac{1}{2^m}\prod_{i=1}^m\Big((a_1^\dagger)^2-(\alpha_i^*+\alpha_{i+m}^*)a_1^\dagger+\alpha_{i}^*\alpha_{i+m}^*\Big)\ket{\mathrm{vac}}  \propto \prod_{i=1}^{2m}(a_1^\dagger-\alpha_i^*)\ket{\mathrm{vac}}\propto\ket{\Psi_{2m,\mathrm{goal}}}\,.
\end{align}

Now, it is important to check the conditions under which the probability of doubling the excitations is not exponentially small. It is easy to show from the previous equations that he probability of success in each step of the doubling is lower bounded by
\begin{align}
 \label{eq:p2mgen2}
 d_{m}\gtrsim \frac{1}{\sqrt{2\pi m}} |\braket{m}{\Psi_{a_1,m}}|^2 |\braket{m}{\Psi_{a_2,m}}|^2\,,
\end{align}
such that it does not decay exponentially if the contribution $|\braket{m}{\Psi_{a_{1,2},m}}|^2$ do not decay exponentially either. This is a reasonable assumption as if $|\braket{m}{\Psi_{a_{1,2},m}}|$ would be exponentially small we could approximate the photonic state without doubling with an exponentially small error. Both Eqs. \ref{eq:gene} and \ref{eq:p2mgen2} show that we can create any arbitrary state with a non-exponentially small probability. Using that $|\braket{m}{\Psi_{a_{1,2},m}}|^2\propto 1/m^\alpha$, it is then easy to show the superpolynomial scaling of probability:
\begin{equation}
R_m \lesssim \frac{m^{\alpha(\log_2 m)+1 }\log_2 m}{2 p}\,.
\end{equation}

\subsection{Increasing the number of excitations with number resolved detection.}

In this last Section, we show how by using atomic number resolved detection, we can overcome the superpolynomial scaling of the previous section. Instead of starting the process again when we detect some excitation in the $a_2$, we can think of using some of these states that still have a non-negligible number of excitations in the $a_1$ mode that we can use a posteriori. To further explore this possibility, we generalize the operation of Eq.~\ref{eqSN:outmD} to see what is the resulting state after $50/50$ beam splitter transformation when we want to sum up $m$ and $n$ excitations in the $a_{1,2}$ modes when we detect the $p$ excitations in the $a_2$ mode:
\begin{align}
\label{eq:outm}
f_p(m,n)&= _1\langle m+n-p\ket{\tilde{\phi}_{m+n-p}}=_1\bra{ m+n-p}\mathbb{P}_{p_2} B \frac{(a_1^\dagger)^m (a_2^\dagger)^n}{\sqrt{m!}\sqrt{n!}}\ket{\mathrm{vac}}\nonumber \\
&=\sum_{k=0}^m  \frac{(-1)^{m-k}}{2^{(m+n)/2}\sqrt{m!}\sqrt{n!}}\binom{m}{k}\binom{n}{m+n-k-p}\sqrt{p!}\sqrt{(m+n-p)!}\,.
\end{align}

\begin{figure}[tb]
	\centering
	\includegraphics[width=0.79\textwidth]{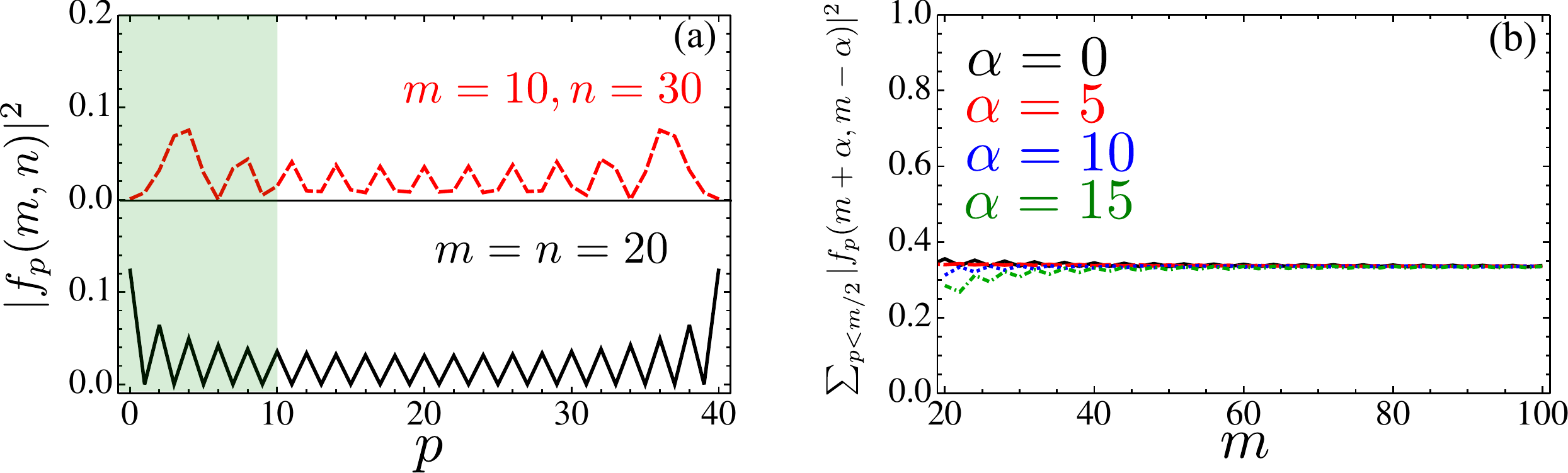}
	\caption{(a) Distribution of probabilities $|f_p(m,n)|^2$ as a function of $p$ for $m=n=20$ (lower panel) and $m=10,n=30$. (b) $\sum_{p<m/2}|f_p(m+\alpha,m-\alpha)|^2$ as a function of $m$ for different degrees of asymmetry given by $\alpha$ as shown in the legend. It can be easily checked that all the cases converges to a constant $s\approx 1/3$. }\label{figSM11}
\end{figure}

In Fig.~\ref{figSM11}(a) we plot the distribution of probabilities $|f_p(m,n)|^2$ as a function of $p$ for different $m$ and $n$. In solid black, we depict the the case with $m=n$ which is the symmetric situation that we consider before. For this symmetric situation the expression of Eq.~\ref{eq:outm} can be simplified as follows:
\begin{align}
\label{eq:outm2}
|f_{2n}(m,m)|^2&= \frac{(2m-2n)! (2n)!}{2^{2m}(m!)^2}\binom{m}{n}^2\approx \frac{1}{\pi}\frac{1}{\sqrt{n(m-n)}}\,.
\end{align}
and $f_{2n-1}(m,m)\equiv 0$ for $n\in \mathbb{N}$, and where in the last approximation we assumed $m,n\gg 1$ to use Stirling approximation. By integrating this expression it can be obtained that the probability of detecting an state $p<m/2$ lead to:
\begin{equation}
\label{eqsm:threehalf}
s_{p<m/2}=\sum_{p=0}^{m/2}|f_p(m,m)|^2\approx \int_0^{m/4} d n|f_{2n}(m,m)|^2=\int_0^{m/4} d n \frac{1}{\pi}\frac{1}{\sqrt{n(m-n)}}= 1/3\equiv s\,,
\end{equation}
which means that there is a probability $s$ independent of $m$ of going from $\ket{m}_1\otimes\ket{m}_2\rightarrow \ket{2m-p>3m/2}_1\otimes\ket{p}_2$. In order to make a worst case estimation, we assume that in each step we go only to $\ket{m}_1\otimes\ket{m}_2\rightarrow \ket{3m/2}_1$. This in principle can be done because if one obtains a higher excitation number it can be decreased by applying infinitesimal beam splitter transformations with an empty mode as:
\begin{align}
\label{eq:e}
\mathbb{P}_{0_2} B(\theta\ll 1)\ket{n}_1\otimes\ket{0}_2 &\approx \left(1-\frac{\theta^2 n}{2}\right)\ket{n}\,,\nonumber \\
\mathbb{P}_{0_2} B(\theta\ll 1)\ket{n}_1\otimes\ket{0}_2& \approx i \theta \sqrt{n} \ket{n-1}_1\,,\nonumber \\
\end{align}
which show that by switching $\theta^2 n\ll 1$, we do not alter $\ket{n}_1$, until we decrease one excitation. This can be applied several times until we arrive to $\ket{3m/2}$. In this worst case scenario, starting again with $M$ single excitations [see Fig.~\ref{figSM10}(c)] we will arrive at least to $m\ge \big(\frac{3}{2}\big)^{\log_2 (M)}$ [which lead $\log_{3/2} m=\log_2 M$] in $\log_2(M)$ steps, which means that with this protocol
\begin{equation}
R_m \lesssim  \frac{\log_2 (M) 2^{\log_2 M-1}}{p s^{\log_2 M}}\approx  \frac{m^{4.41}\log_{3/2} m}{2 p}\,,
\end{equation}
where we used $\log_b x=\frac{\log_a x}{\log_a b}$. Thus, by using number resolved detection we turn the superpolynomial scaling into polynomial which is big improvement if we want to scale our protocol for larger excitation numbers. In practical situations, the scaling will be better as one will never decrease the excitations to arrive to $3n/2$, but will use the larger excitation numbers. The consequence of that is that within the tree we will need to add up excitation in an asymmetric way, i.e., $(m,n)$ as we showed in Fig. \ref{figSM11}(b). However, even in this case it can also be shown that 
\begin{equation}
\label{eqsm:threehalf}
\sum_{p=1}^{m/2}|f_p(m+\alpha,m-\alpha)|^2\approx 1/3\equiv s\,,
\end{equation}
for approximately all $\alpha$ as shown numerically in Fig. \ref{figSM11}(b). This proves that the number resolved excitation also helps even in the asymmetric configuration as it also may give rise to a probability in each step independent of the number of excitations we add up.

 \subsection{Atomic number resolved detection using waveguide QED.}

 Though standard linear optics setups can already distinguish between photon numbers using superconducting detectors, current technologies are far from obtaining the desired detection efficiencies for optical photons. In our case, we can benefit directly from our atomic detection which naturally give rise to ways of having number resolved detection. In this Section, we make a short discussion of how one can do it.
 
\begin{figure}[tb]
	\centering
	\includegraphics[width=0.69\textwidth]{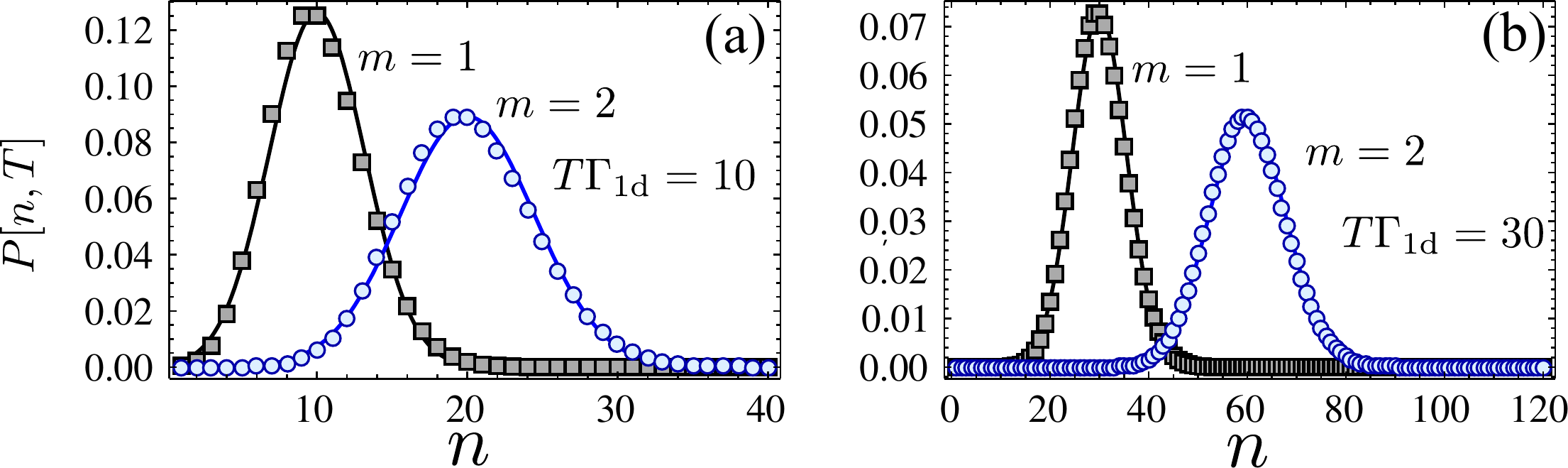}
	\caption{(a-b) Photon counting probabilities $P[n,T]$ for $T\Gamma_\oned=10-30$ for two different atomic levels with $m=1$ [black] and $m=2$ excitations. }\label{figSM12}
\end{figure}

 The goal is to be able to distinguish in one atomic level between having $m$ or $m+1$ excitations. In order to do it, we can drive a closed atomic transition to an excited state such that we emit photons through the waveguide. Using the fact that $P_\oned \gg 1$, we will be able to obtain the photon counting distribution $P[n,T]$ of having detected $n$ photons after a time interval $T$. Interestingly, this photon counting distribution is given by a Poisson process with different parameter $\lambda_m=m\Gamma_\oned T$, being $m$ the number of excitations, i.e., $P[n,T]=\frac{\lambda_m^n}{n!} e^{-\lambda_m}$. In Fig.~\ref{figSM12}, we see this distribution quickly converges to a Gaussian with mean value $\lambda_m$ and width $\sqrt{\lambda_m}$, such that for large $T$ (or equivalently large $\lambda_m$), the overlap between the two distribution, $\varepsilon$ vanishes approximately with $T$ approximately $\varepsilon\propto e^{-(\Gamma_\oned T)^2}$, which give as some bounds of the $T$ required to minimize the error introduced in the number resolved detection.
 
 Depending on the particular implementation, a more thorough analysis of the errors has to be performed, to estimate the final fidelities of the protocol, though our paper serves as an optimistic basis for future research.

\end{document}